\documentclass[aps,prb,floatfix,amsmath,amssymb,preprint,eqsecnum,nofootinbib]{revtex4}
\usepackage{graphicx}
\usepackage{dcolumn}
\usepackage{bbm}
\usepackage{epstopdf}
\usepackage{color}

\usepackage{setspace} 
\newcommand{\beq}{\begin{equation}}
\newcommand{\eeq}{\end{equation}}
\newcommand{\beqn}{\begin{eqnarray}}
\newcommand{\eeqn}{\end{eqnarray}}
\newcommand{\nn}{\nonumber\\}

\usepackage{amsmath}
\usepackage{amssymb}

\def \v{{\mathbf{v}}}
\def \u{{\mathbf{u}}}
\def \q{{\mathbf{q}}}
\def \R{{\mathbf{R}}}
\def \Q{{\mathbf{Q}}}

\begin{document}

\title{Vison states and confinement transitions\\ of $\mathbb{Z}_2$ spin liquids on the kagome lattice}

\author{Yejin Huh}
\affiliation{Department of Physics, Harvard University, Cambridge MA
02138}

\author{Matthias Punk}
\affiliation{Department of Physics, Harvard University, Cambridge MA
02138}

\author{Subir Sachdev}
\affiliation{Department of Physics, Harvard University, Cambridge MA
02138}

\date{\today \\
\vspace{1.6in}}

\begin{abstract}We present a projective symmetry group (PSG) analysis of the spinless excitations of $\mathbb{Z}_2$ spin
liquids on the kagome lattice. In the simplest case, vortices carrying $\mathbb{Z}_2$ magnetic flux (`visons') are shown to transform
under the 48 element group ${\rm GL}(2,\mathbb{Z}_3)$. Alternative exchange couplings can also lead to a second
case with visons transforming under 288 element group ${\rm GL}(2,\mathbb{Z}_3) \times {\rm D}_3$.
We study the quantum phase transition in which visons condense into
confining states with valence bond solid order. The critical field theories 
and confining states are classified
using the vison PSGs.
\end{abstract}

\maketitle

\section{Introduction}
\label{sec:intro}

Frustrated quantum magnets have a long history in condensed matter physics. Due to their relative simplicity in comparison to itinerant electron systems, they provide an ideal playground to study strongly correlated states of matter and the effects of competing ground states. One long standing goal is to identify realistic systems that realize spin liquid ground states, \emph{i.e.}~strongly correlated states of localized spins that don't break any symmetries. A promising candidate is the antiferromagnetic spin-1/2 Heisenberg model on the kagome lattice and its realization in nature in the from of the mineral Herbertsmithite. \cite{herbertsmithite,YLee} Until recently several theoretical works based on dimer model approaches, series expansions as well as numerical calculations suggested that the ground state of this model is not a spin liquid, but a valence bond solid \cite{kag1,nikolic,kag3,kag4,kag5}, \emph{i.e.} a state that doesn't break the spin rotation symmetry, but instead breaks a lattice symmetry. 
However, a recent DMRG study of Yan {\em et al.} \cite{white} has provided striking evidence for a spin liquid
ground state for the $S=1/2$ Heisenberg antiferromagnet on the kagome lattice. Yan {\em et al.} found
a gap to all excitations, and it is plausible that their ground state realizes a $\mathbb{Z}_2$ spin liquid. \cite{rs2,wen1,sskag,wv,mot,rankag,didier1,didier2}

The results of Yan {\em et al.} also indicate the presence of proximate valence bond solid (VBS) states in which
the space group symmetry of the kagome lattice is broken, and the fractionalized excitations of the spin liquid
are confined into integer spin states. 
The confinement quantum phase transition should be accessible in extended models with further neighbor exchange interactions,
and numerical studies of such transitions can serve as a valuable probe of characteristics of the spin liquid.

This paper shall classify elementary vortex excitations of the $\mathbb{Z}_2$ spin liquid, carrying 
$\mathbb{Z}_2$ magnetic flux\cite{readc,rs2,sf},
often called `visons', which are analogous to the Abrikosov vortices of BCS superconductors\cite{ijmpb} (after electromagnetism is replaced by 
strong coupling to a compact U(1) gauge theory).
We will compute
their projective symmetry group\cite{wenpsg} (PSG), and their spectrum using an effective frustrated Ising model.\cite{jalabert}
Note that the Ising `spin' has nothing to do with the $S=1/2$ spin of the underlying antiferromagnet,
and it is instead the creation or annihliation operator of the vortex excitation which is centered on sites of a lattice
dual to that of the antiferromagnet.
For the kagome antiferromagnet, the Ising model resides on the dice lattice, and the simplest effective Ising model
has a degenerate momentum-independent 
spectrum.\cite{nikolic,ks} We shall show how the PSG constraints allow a systematic analysis of further 
neighbor interactions in the effective Ising model. Such extended interactions must generically be present,\cite{xu2} and 
they lead to well-defined vison states
with a finite effective mass. Depending upon the values of these effective interactions, we find 2 possibilities for the vison states:
the simplest case has them transforming under the 48 element group ${\rm GL}(2,\mathbb{Z}_3)$, the group of $2 \times 2$ matrices
with non-zero determinant whose matrix elements belong to the field $\mathbb{Z}_3$. The more complex case has 
visons states of the 288 element group ${\rm GL}(2,\mathbb{Z}_3) \times {\rm D}_3$.

\begin{figure}
\begin{center}
\includegraphics[height=5cm]{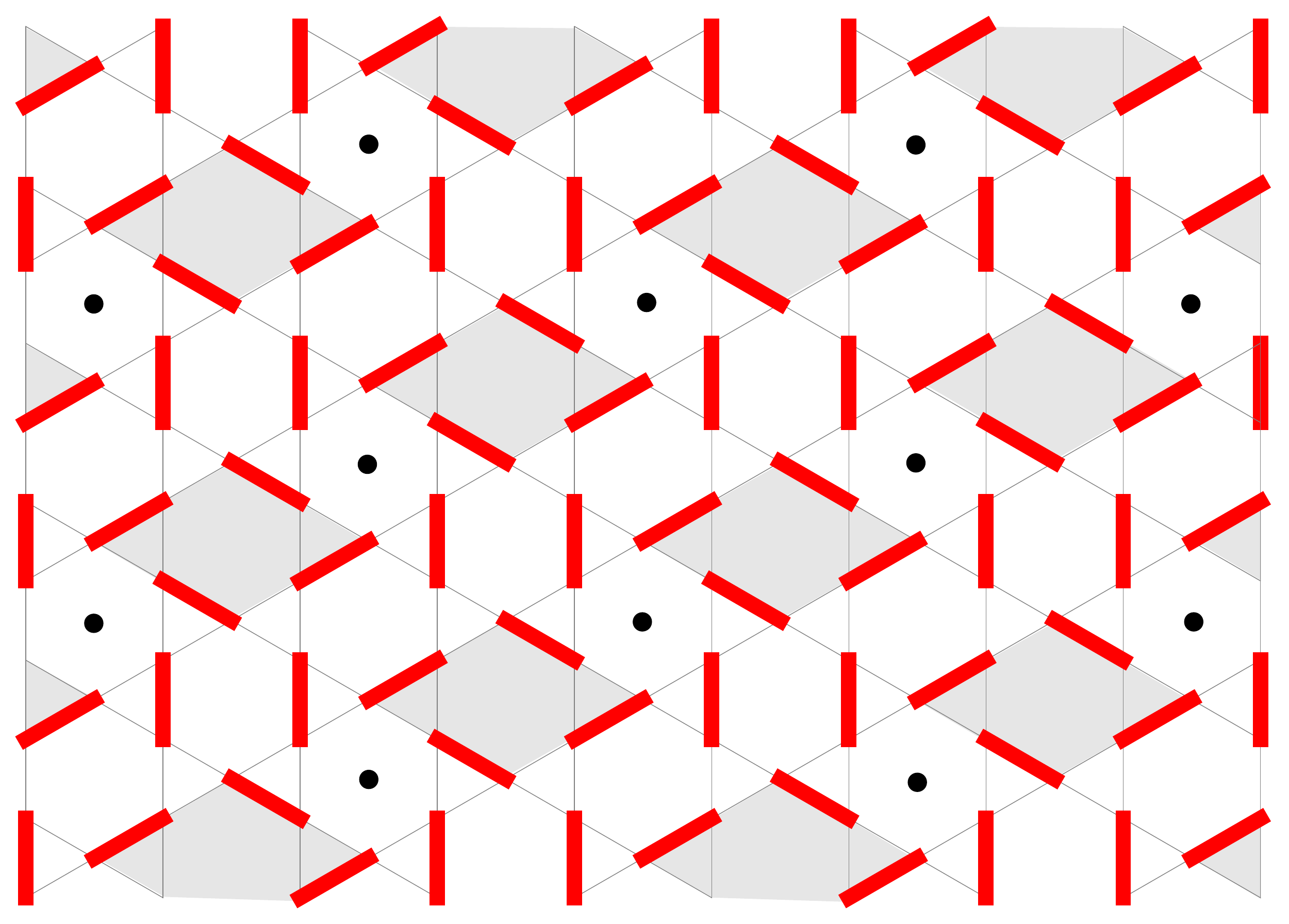}
\hspace{1cm}
\includegraphics[height=5cm]{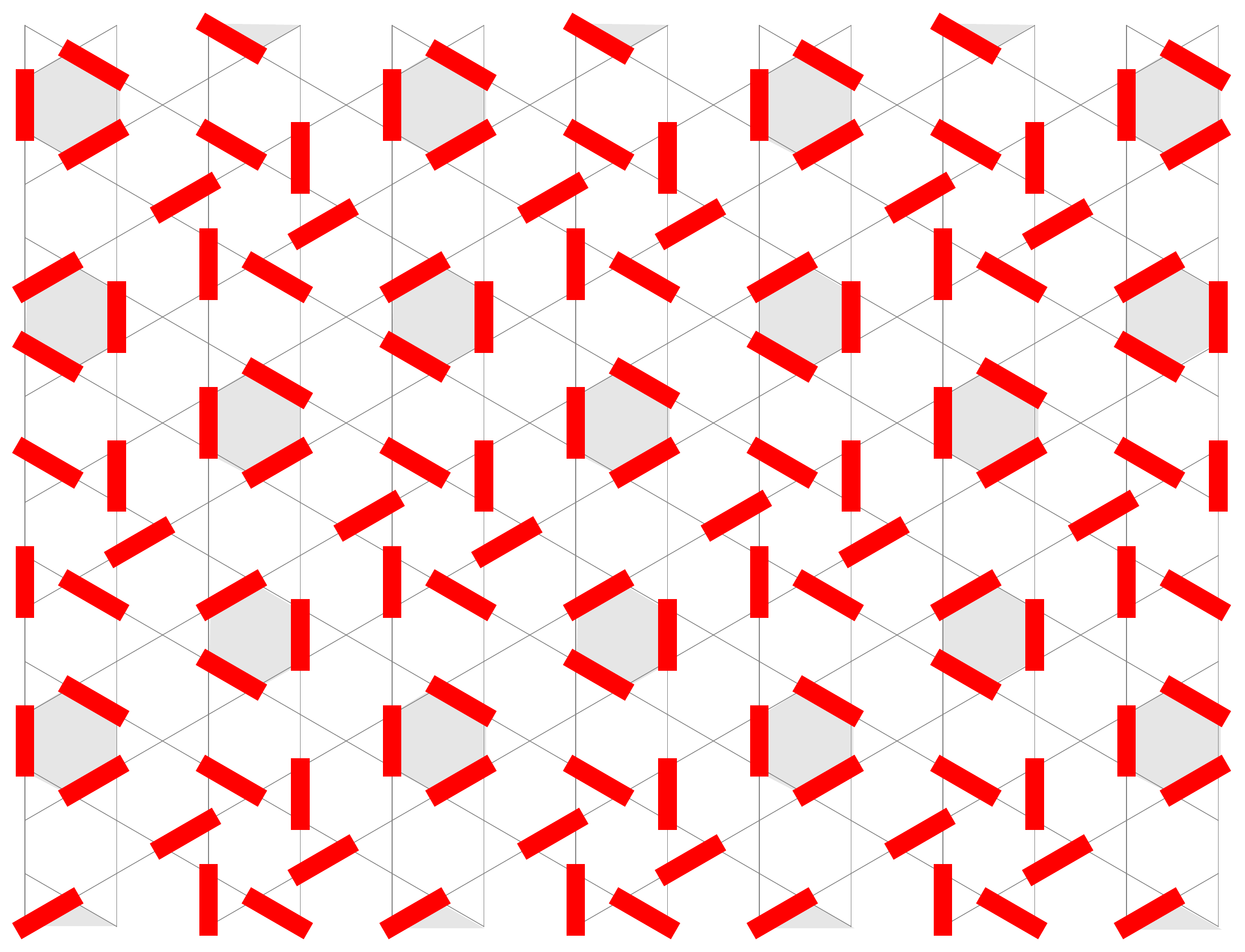}
\end{center}
\caption{Visualization of two different valence bond solid states as dimer coverings of the kagome lattice. Each dimer represents a frustrated bond in the ordered phase of the corresponding Ising model on the dual dice lattice. The left pattern, representing the VBS $1_\text{F}$ phase (see text), is a hardcore dimer covering with a 12-site unit cell that maximizes the number of perfectly flippable diamonds (highlighted in gray). Note however, that the choice of diamonds is not unique because the dimer pattern is symmetric with respect to $\pi/3$-rotations around the hexagons marked by a black dot. The pattern on the right represents the VBS $1_\text{A}$ phase (see text), has a 36-site unit cell and maximizes the number of perfectly flippable hexagons (highlighted in gray).}
\label{figkagome}
\end{figure}

Armed with this description of the vison states, will 
propose quantum field theories for the confinement transitions of $\mathbb{Z}_2$ spin liquids
on the kagome lattice. These transitions are associated with condensation of visons,\cite{sf,jalabert,vojta,sondhi,xu,xu2,didier3} and are expressed in 
terms of a multi-component `relativistic' scalar field; the field theory with ${\rm GL}(2,\mathbb{Z}_3)$ symmetry appears in
Eq.~(\ref{GL1}). These field theories also place constraints on the specific patterns of spatial broken symmetry in the 
confining VBS states found next to the quantum critical point and we will present phase diagrams illustrating these states. Visualizations of two possible VBS states are shown in Fig.\ \ref{figkagome}.
The left VBS pattern in Fig.~\ref{figkagome} appears for the simplest case of the ${\rm GL}(2,\mathbb{Z}_3)$ visons,
and it is interesting that it is closely related to the ``diamond pattern'' which is enhanced in the numerical studies of 
Yan {\em et al.}\cite{white}. The right pattern in Fig.~\ref{figkagome} is one of many possible VBS states for 
the ${\rm GL}(2,\mathbb{Z}_3) \times {\rm D}_3$ visons, and maps to the ``honeycomb'' VBS states found
in earlier studies \cite{kag1,kag3,kag4,kag5,nikolic}.

We will begin in Section~\ref{sec:visons} with a review of the basic characteristics of $\mathbb{Z}_2$ spin liquids and of their vison excitations.
A key property of a vison is that it picks up a Aharanov-Bohm phase of $\pi$ upon encircling every $S=1/2$ spin on the sites of the 
antiferromagnet;\cite{jalabert,sf,vojta,sondhi}, as will be described in Section~\ref{sec:visons}. 

Section~\ref{sec:dice} contains our main new results. We begin with general effective theories of vison motion which incorporate the Aharanov-Bohm phase\cite{xu,xu2} of $\pi$: for the kagome antiferromagnet, these are conveniently expressed in terms of an effective, fully-frustrated
Ising model on the dice lattice. A key parameter in this Ising model is the sign of a particular next-nearest-neighbor interaction.
A `ferromagnetic' sign leads to the simpler vison PSG, and is discussed in Section~\ref{sec:ferro};
the more complicated `antiferromagnetic' case is discussed in Section~\ref{sec:antiferro}.
In both cases, we use the vison PSG to present quantum field theories for the confinement transitions, and
discuss renormalization group analyses of these quantum critical points. The field theories are also used to classify the patterns
of lattice symmetry breaking in the confining valence bond solid states.

In Appendix \ref{app:berry} the explicit calculation of a visons Berry phase can be found. The results of a PSG analysis of visons for different lattice geometries are summarized in Appendix \ref{app:geom}.

\section{$\mathbb{Z}_2$ spin liquids and visons}
\label{sec:visons}

We begin by a review of the basic properties of $\mathbb{Z}_2$ spin liquids, following the description in Ref.~\onlinecite{sskag}.

It is convenient to describe the $S=1/2$ spins $\vec{S}_i$ using Schwinger bosons\cite{arovas} $b_{i \alpha}$ ($\alpha = \uparrow, \downarrow$)
\beq
\vec{S}_i = \frac{1}{2} b_{i \alpha}^\dagger \vec{\sigma}_{\alpha\beta} b_{i \beta}, 
\eeq
where $\vec{\sigma}$ are the Pauli matrices, and the bosons obey the local constraint
\beq
\sum_\alpha b_{i \alpha}^\dagger b_{i \alpha} = 1 \label{const}
\eeq
on every site $i$. Our analysis below can be easily extended to gapped $\mathbb{Z}_2$ spin liquids obtained from the Schwinger fermion
formulation,\cite{wen1,rankag,didier1,didier2} but we will only consider the Schwinger boson case for brevity.

The $\mathbb{Z}_2$ spin liquid is described by an effective boson Hamiltonian
\beq
\mathcal{H}_b =  - \sum_{i < j} Q_{ij} \varepsilon_{\alpha\beta} b_{i \alpha}^\dagger b_{j \beta}^\dagger + \mbox{H.c.} + \lambda \sum_i b_{i \alpha}^\dagger
b_{i \alpha}, \label{Hb}
\eeq
where $\varepsilon$ is the antisymmetric unit tensor, $\lambda$ is chosen to satisfy the constraint in Eq.~(\ref{const}) on average, 
and the $Q_{ij}=-Q_{ji}$ are a set of variational parameters chosen to optimize the energy of the spin liquid state. Generally, the $Q_{ij}$ are
chosen to be non-zero only on near neighbor links. The `$\mathbb{Z}_2$' character of the spin liquid requires that the links with non-zero $Q_{ij}$
can form closed loops with an odd number of links,\cite{rs2,wen1,sskag}. If the state preserves time-reversal symmetry, all the $Q_{ij}$ 
can be chosen real.

This effective Hamiltonian also motivates a wavefunction of the spin liquid\cite{rs1,mot}
\beq
\left|\mbox{SL} \right\rangle = \mathcal{P} \exp \left( \sum_{i<j} f_{ij} \, \varepsilon_{\alpha\beta} b_{i \alpha}^\dagger b_{j \beta}^\dagger
\right) |0 \rangle, \label{spinl}
\eeq
where $|0 \rangle$ is the boson vaccum, $\mathcal{P}$ is a projection operator which selects only states which obey Eq.~(\ref{const}), 
and the boson pair wavefunction $f_{ij}=-f_{ji}$ is determined by diagonalizing Eq.~(\ref{Hb}) by a Bogoliubov transformation
(see Eq.~(\ref{fij}) in Appendix~\ref{app:berry}).

\begin{figure}
\begin{center}
\includegraphics[width=3.5in]{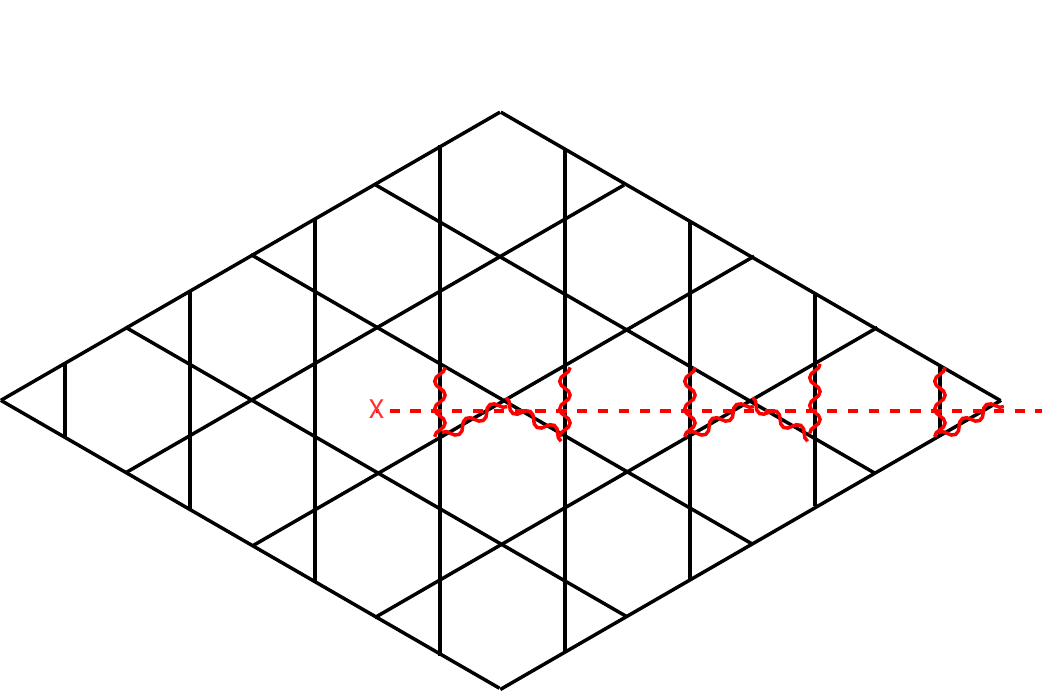}
\end{center}
\caption{A vison on the kagome lattice. The center of the vison is marked by the X. We have $\mbox{sgn}(Q^{v}_{ij}) = - \mbox{sgn}(Q_{ij})$
only on the links marked by the wavy lines.}
\label{branchcut}
\end{figure}

The Schwinger boson approach also allows a description of the vison excited states.\cite{rs2,ijmpb} 
We choose the vison state as the ground state of a Hamiltonian, $\mathcal{H}_b^v$, obtained from $\mathcal{H}$ by mapping
$Q_{ij} \rightarrow Q^v_{ij}$ (see Eq.~(\ref{Ht})); then the vison state $|\Psi^v \rangle$ has a wavefunction as in Eq.~(\ref{spinl}),
but with $f_{ij} \rightarrow f^v_{ij}$ (see Eq.~(\ref{vpsi})). 
Far from the center of the vison, we have $|Q^{v}_{ij}| = |Q_{ij}|$, while closer to the center there
are differences in the magnitudes. However, the key difference is in the signs of the  link variables, as illustrated in Fig.~\ref{branchcut}:
there is a `branch-cut' emerging from the vison core along which $\mbox{sgn}(Q^{v}_{ij}) = - \mbox{sgn}(Q_{ij})$. 
This branch-cut
ensures that the $\mathbb{Z}_2$ flux equals -1 on all loops which encircle the vison core, while other loops do not have non-trivial $\mathbb{Z}_2$ flux. 
Such a configuration of the $Q^v_{ij}$ is expected to be a 
metastable solution of the Schwinger boson mean-field equations, representing a vison excitation.

\begin{figure}
\begin{center}
\includegraphics[width=3.0in]{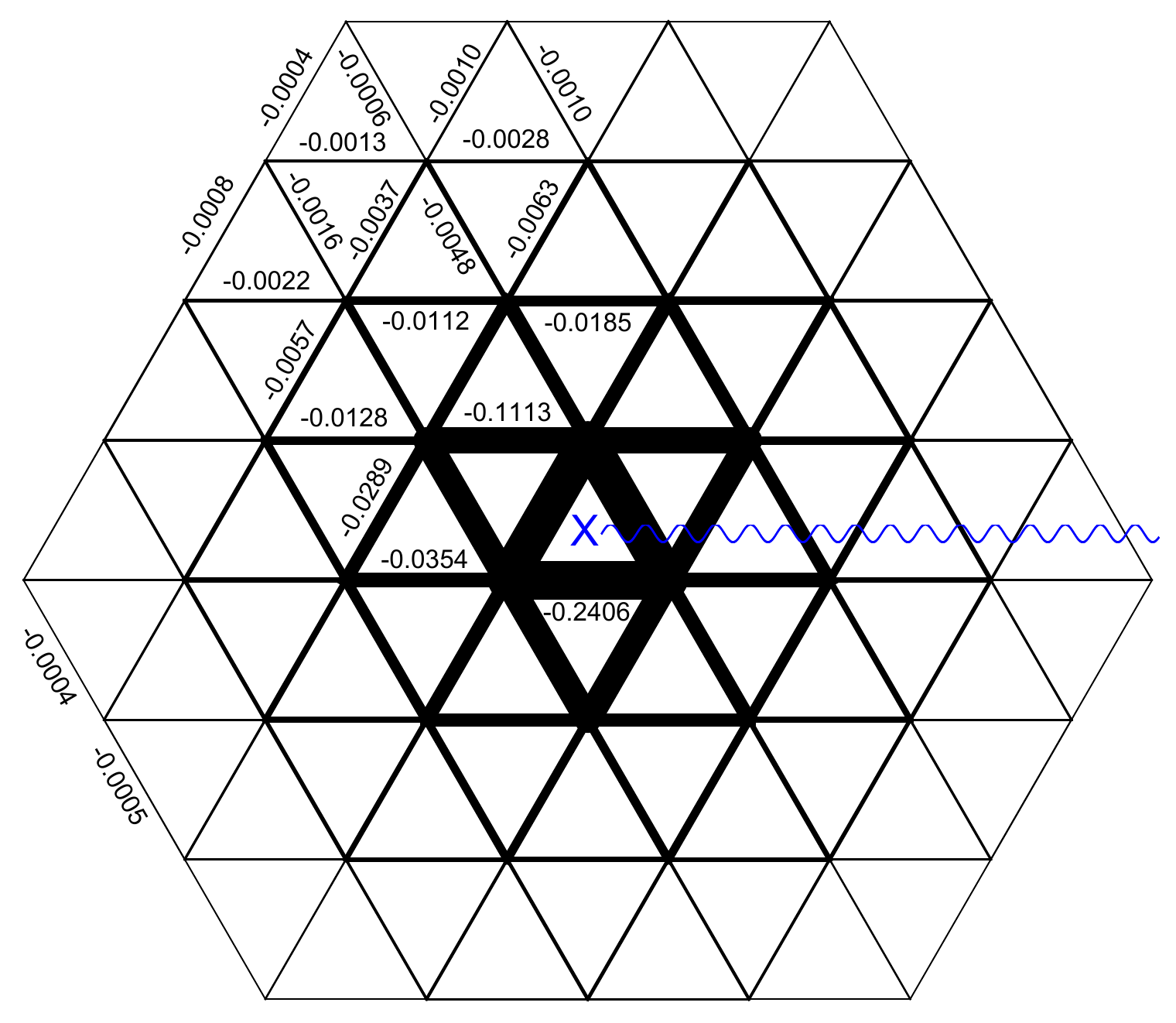}
\end{center}
\caption{A vison on the triangular lattice (a similar solution is expected on the kagome lattice). The center of the vison is marked by the X. The wavy line is the `branch-cut' where we have $\mbox{sgn}(Q^{v}_{ij}) = - \mbox{sgn}(Q_{ij})$ only on the links crossed by the line. Plotted is the minimization result of $E[ \{ Q_{ij} \}]$ with $\alpha=1, \beta=-2, K=-0.5$. Minimization is done with the cluster embedded in a vison-free lattice with all nearest neighbor links $Q_{ij}$. The numbers are $(Q_{ij}-Q^{v}_{ij})$ and the thickness of the links are proportional to $(Q^{v}_{ij}-Q_{ij})^{1/2}$. $K<0$ gives rise to the zero flux state while $K>0$ favors the $\pi$ flux state where the unit cell is doubled. The zero and $\pi$ flux states without the vison have been studied previously.\cite{sskag, wv2} }  
\label{vison}
\end{figure}

The previous solution for the vison state \cite{ijmpb} was obtained using a continuum field theoretic representation of the Schwinger
boson theory, valid in the limit of a {\em small\/} energy gap towards spinful excitations. In principle, such a solution can also be obtained
by a complete solution of the Schwinger boson equations on the lattice, but this requires considerable numerical effort. Here, we illustrate
the solution obtained in the of a {\em large\/} spin gap. In the large spin gap limit, \cite{tchern} we can integrate out the Schwinger bosons,
and write the energy as a local functional of the $Q_{ij}$. This functional is strongly constrained by gauge-invariance: for time-independent
$Q_{ij}$, this functional takes the form
\beq
E[\{Q_{ij}\}] = - \sum_{i<j} \left( \alpha |Q_{ij}|^2 + \frac{\beta}{2} |Q_{ij} |^4 \right)  + K \sum_{{\rm even\, loops}} Q_{ij} Q_{jk}^\ast 
\ldots Q_{\ell i}^\ast
\eeq
Here $\alpha$, $\beta$, and $K$ are coupling constants determined by the parameters in the Hamiltonian of the antiferromagnet.
We have shown them to be site-independent, 
because we have only displayed terms in which all links/loops are equivalent; they can depend upon links/loops
for longer range couplings 
provided 
the full lattice symmetry is preserved.
We describe the results of a minimization
of $E[ \{ Q_{ij} \}]$ on the simpler case of the triangular lattice in Fig.~\ref{vison}. 
The magnitudes of $Q^v_{ij}$ are suppressed close to the vison, and converge to $Q_{ij}$ as we move away from the vison (modulo the sign change associated with the branch cut), analogous to the Abrikisov vortices. Despite the branchcut breaking the 3-fold rotation symmetry, 
the gauge-invariant fluxes of $Q^v_{ij}$ preserve the rotation symmetry. 

\begin{figure}
\begin{center}
\includegraphics[width=2in]{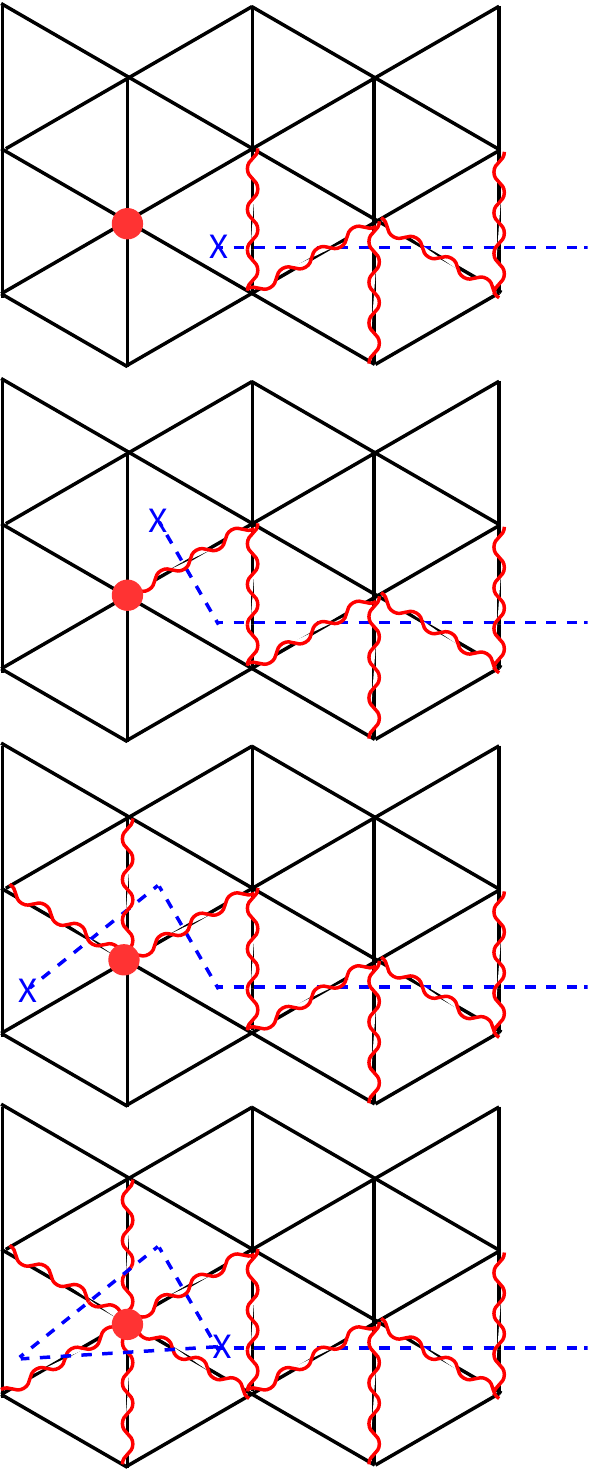}
\end{center}
\caption{Periodic motion of a vison around a closed loop $\mathcal{C}$ on the triangular lattice. 
Here $\mathcal{C}$ encloses the single site marked by the filled circle.
The wavy lines indicate 
$\mbox{sgn}(Q^{v}_{ij}) = - \mbox{sgn}(Q_{ij})$, as in Fig.~\ref{branchcut}. The bottom state is gauge-equivalent to the top
state, after the gauge transformation $b_{i \alpha} \rightarrow - b_{i \alpha}$ only for the site $i$ marked by the filled circle.}
\label{visonberry}
\end{figure}

Let us now consider the motion of a single vison. A key ingredient is the Berry phase a vison accumulates while moving through the background
spin liquid.
The gauge-invariant Berry phases are those associated with a periodic motion, and so let us consider the motion of a vison along a general
closed loop $\mathcal{C}$. We illustrate the simple case where $\mathcal{C}$ encloses a single site of the triangular lattice antiferromagnet
in Fig.~\ref{visonberry}.
The Berry phase for this periodic motion can be computed in a manner
analogous to that presented in Section III.A of Ref.~\onlinecite{rsb} for a monopole in a U(1) spin liquid; details appear in 
Appendix~\ref{app:berry}. The final gauge-invariant Berry phase turns out to be
given by the gauge-transformation required to map the final state to the initial state. The analysis in Fig.~\ref{visonberry} shows that
the required gauge transformation is
\beqn
b_{i \alpha} \rightarrow - b_{i \alpha}, &\quad& \mbox{for $i$ inside $\mathcal{C}$} \nn
b_{i \alpha} \rightarrow b_{i \alpha}, &\quad& \mbox{for $i$ outside $\mathcal{C}$}.
\eeqn
By Eq.~(\ref{const}), each site has one boson, and so the total Berry phase accumulated by $|\Psi^v \rangle$ is 
\beq
\pi \times \left( \mbox{number of sites enclosed by $\mathcal{C}$} \right),
\eeq
as recognized in earlier works \cite{sf,jalabert,vojta,xu,xu2}. It is also clear that for a spin $S$ antiferromagnet, the Berry phase would be multiplied
by a factor of $2S$.

Finally, let us also mention the spinon states, although these will not play a role in the subsequent analysis of the present paper.
These are created by applying the Bogoliubov quasiparticle operator $\gamma_{\mu \alpha}^\dagger$ (Appendix~\ref{app:berry})
on the spin-liquid ground state; in this manner we obtain the spinon state
\beqn
|\mu \alpha \rangle &=&  \sum_{\ell} \left(U^{-1 \ast} \right)_{\mu \ell } | \ell \alpha \rangle \nn
| \ell \alpha \rangle &=& \mathcal{P}\, b^\dagger_{\ell \alpha} 
\exp \left( \sum_{i<j} f_{ij} \, \varepsilon_{\alpha\beta} b_{i \alpha}^\dagger b_{j \beta}^\dagger
\right) |0 \rangle, \label{spinon}
\eeqn
where $U_{i \mu}$ is a Bogoliubov rotation matrix defined in Appendix~\ref{app:berry}.
We can now also consider a spinon well-separated from a vison, and describe the motion of a vison along a large contour $\mathcal{C}$ which
encircles the spinon. First, consider the motion of a vison around the spinon state $| \ell \alpha \rangle$ localized on the site $\ell$. Proceeding with the 
argument as above, the projection onto states which obey Eq.~(\ref{const}) now implies that the Berry phase for such a process
is $\pi \times (( \mbox{number of sites enclosed by $\mathcal{C}$}) - 1)$. The transformation to the spinon state $| \mu \alpha \rangle$ will
not change the result, provided the `wavefunction' $\left(U^{-1 \ast} \right)_{\mu \ell }$ is localized well within the contour $\mathcal{C}$. 
Thus relative to the Berry phase in the case without a spinon, the vison acquires an additional phase of $\pi$ upon encircling a spinon
{\em i.e.\/} the spinons and visons are relative semions.\cite{readc}

\section{PSG analysis of the fully frustrated Ising model on the dice lattice}
\label{sec:dice}

The analysis of Section~\ref{sec:visons} suggests a simple effective model for the vison fluctuations about the 
$\mathbb{Z}_2$ spin liquid ground state. In the framework of the path-integral formations of Ref.~\onlinecite{sskag},
the visons are saddle-points of the $Q_{ij}$ with $\pi$-flux, as shown in Fig.~\ref{vison}.
Each saddle-point traces a world-line in spacetime, representing the time evolution of the vison.
The Berry phase computation in Section~\ref{sec:visons} shows that this world-line picks up $\pi$-flux
each time it encircles a kagome lattice site. We know further that two such worldlines can annihilate each other,
because $2 \pi$ magnetic flux is equivalent to zero flux. So if we interpret the wordlines as the trajectory of a particle,
that particle must be its own anti-particle, and has a real field operator. In this manner, we see that the fluctuations represented
as the sum over all vison worldlines is precisely that of a frustrated Ising model in a transverse field \cite{jalabert,vojta,sf}:
\beq
H = -  \sum_{i<j} J_{i j} \, \phi_i \phi_j  + \ldots \  ,
\label{HIsing}
\eeq
where the product of bonds around each elementary plaquette is negative
\beq
\prod_{\text{plaq.}} \text{sgn} (J_{ij}) = -1 \  .
\label{frustr}
\eeq
Also, we have not displayed the transverse-field term, because most of our symmetry considerations are restricted to time-independent,
static configurations.

\begin{figure}
  \begin{center}
    \includegraphics[width=4cm]{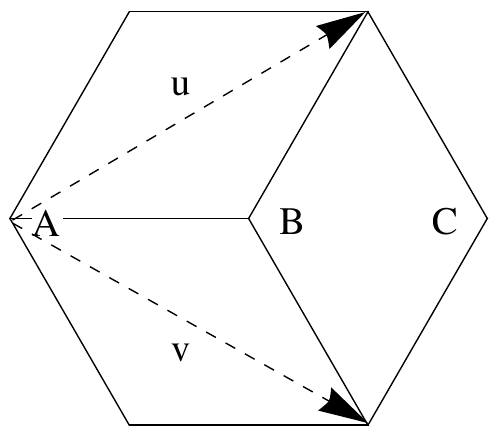}
    \hspace{2cm}
    \includegraphics[width=4cm]{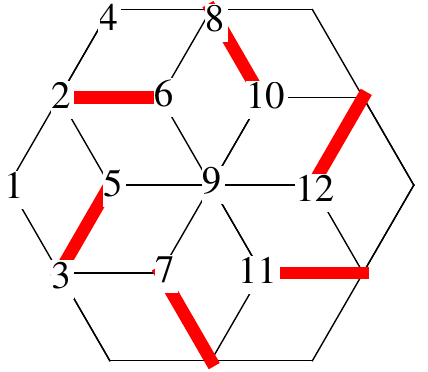}
    \caption{Left: unit cell of the dice lattice consisting of the 3 independent sites labeled $A$, $B$ and $C$. The $A$ sites are 6-coordinated, whereas the $B$ and $C$ sites are 3-coordinated. The basis vectors $\mathbf{u}=(3/2, \sqrt{3}/2)$ and $\mathbf{v}=(3/2,-\sqrt{3}/2)$ are indicated by dashed arrows. Right: gauge choice for the fully frustrated Ising model on the dice lattice with a 12-site unit cell. Red thick bonds are frustrated, {\em i.e.\/} $\text{sgn}(J_{ij})=-1$.}
    \label{fig:dice}
  \end{center}
\end{figure}

In the following we will use a soft-spin formulation where the $\phi_j$'s take real values. This model is invariant under $\mathbb{Z}_2$ gauge transformations $\phi_j \to \sigma_j \phi_j$, $J_{ij} \to \sigma_i \sigma_j J_{ij}$ with $\sigma_j = \pm 1$. For the case of the $\mathbb{Z}_2$ spin liquid on a kagome lattice the dual Ising model lives on the dice lattice, shown in Fig.\ \ref{fig:dice}. The dice lattice has three independent sites per hexagonal unit cell, two of them are three-coordinated and one is six-coordinated. Flipping a dual Ising spin changes the flux through a plaquette on the kagome lattice by $\pi$, thereby creating or annihilating a vison. 
The magnetically disordered phase of the Ising model corresponds to a $\mathbb{Z}_2$ spin liquid with deconfined spinon excitations, whereas the ordered phases describe different valence bond solids where the visons are condensed and fractional excitations are confined.

\begin{figure}
  \begin{center}
    \includegraphics[width=4cm]{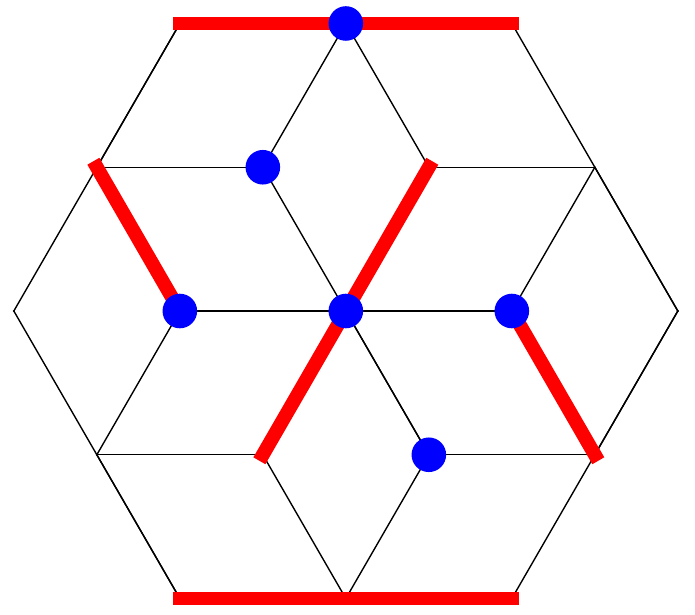}
    \hspace{2cm}
     \includegraphics[width=4cm]{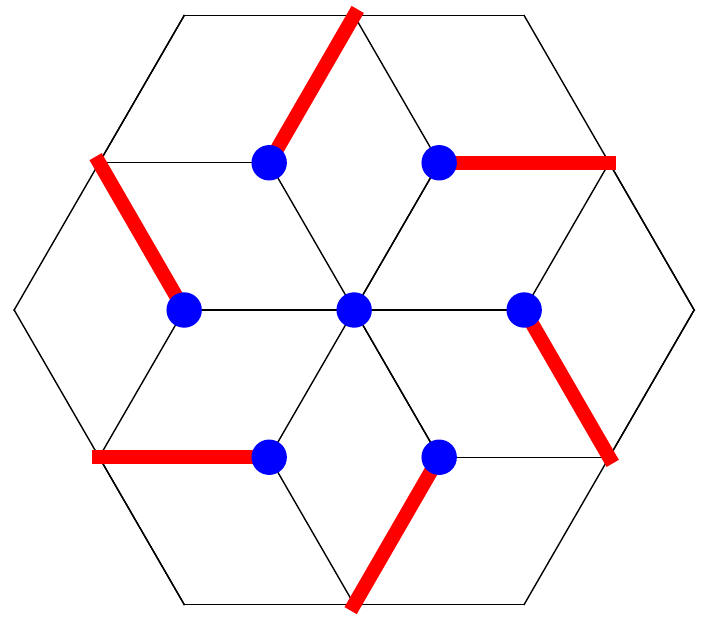}
    \caption{Gauge transformations associated with translations by $\u$ and reflections about the x-axis. Shown is the unit cell of the fully frustrated dice model after a lattice translation by $\u$ (left) and after a reflection about the x-axis (right). The corresponding gauge-transformations consist of spin flips on the sites marked by blue points, which restore the original gauge pattern as in Fig.\ \ref{fig:dice}.}
    \label{fig:PSG}
  \end{center}
\end{figure}

In the following we study the confinement transitions of the spin liquid by constructing a Ginzburg-Landau functional that is consistent with the projective symmetry group (PSG), {\em i.e.\/}  the combination of lattice-symmetry- and $\mathbb{Z}_2$ gauge-transformations that leave the Hamiltonian \eqref{HIsing} together with \eqref{frustr} invariant. In order to determine the PSG transformations we fix the gauge of nearest neighbor interactions as shown in Fig.\ \ref{fig:dice}, thereby obtaining a unit cell with twelve sites. The generators of the dice lattice symmetry group are translations by one of the two basis vectors, {\em e.g.\/} $\u$, reflections about one axis, {\em e.g.\/} the $x$-axis, and $\pi/3$ rotations about the central 
6-coordinated site. Since our gauge choice is already invariant under rotations, we only need to determine the gauge-transformations corresponding to translations and reflections to specify the PSG. These are shown in Fig.\ \ref{fig:PSG}.

\begin{figure}
  \begin{center}
    \includegraphics[width=4cm]{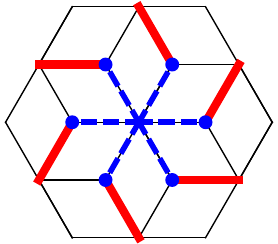}
    \caption{Additional next-nearest neighbor interactions between 3-coordinated sites on the dice lattice (shown as dashed blue lines) that are allowed by the PSG. For illustrative clarity not all additional bonds in the unit cell are shown. All other bonds can be obtained from the ones that are shown via translations by the basis vectors $\u$ and/or $\v$.}
    \label{fig:nnnInt}
  \end{center}
\end{figure}

The dispersion relations of the soft-spin modes can be obtained directly from the Hamiltonian \eqref{HIsing}. The corresponding action takes the form
\beq
S= \sum_{\Omega,\q}  \phi^{(i)}_{\q, \Omega} \big[ (\Omega^2 + m^2) \delta_{i,j} - J^{(ij)}_\q \big] \phi^{(j)}_{-\q, -\Omega}
\eeq
where the summation over the sublattice indices $i,j=1...12$ is implicit, $J^{(ij)}_\q$ denotes the Fourier transform of the interaction matrix $J_{ij}$, and we have included a kinetic energy with frequency $\Omega$ (which descends from the transverse field), and a mass term, $m$. As noted earlier\cite{nikolic,ks}, the frustrated Ising model on the dice lattice with nearest neighbor interactions gives rise to three flat bands (each being four-fold degenerate in our gauge choice with a 12-site unit cell), which would result in infinitely many critical modes. In order to lift this degeneracy we include further interactions beyond nearest neighbors\cite{xu2} that are consistent with the PSG.
Physically, these interactions correspond to the hopping of visons beyond nearest neighbors.
Different masses on the 3- and 6-coordinated sites would be allowed by the PSG, but they don't give rise to a momentum dependence of the modes and thus don't change the picture qualitatively. Out of the five possible additional interactions up to a distance of two times the nearest neighbor bond length only one is consistent with the PSG. This interaction, shown in Fig.\ \ref{fig:nnnInt}, connects different 3-coordinated lattice sites and gives rise to a non-flat dispersion. The Fourier transform of the interaction matrix $J_{ij}$ takes the form
\begin{equation}
\setcounter{MaxMatrixCols}{12}
J^{(ij)}_\mathbf{q}=\left[ \begin{smallmatrix}
0 & 1& 1 & \kappa^*_1  & 0 & 0 & 0 &  e^{-i 2 \q \u} & 0 & 0 & 0 & 0 \\
1 & 0 & 0 & 1 & 1 & -1& 0 & 0 & 0 & 0 & -e^{-i 2 \q \v} & e^{-i 2 \q \v} \\
1 & 0 & 0 & e^{i 2 \q (\v-\u)} & -1 & 0 & 1& 0 & 0 & e^{-i 2 \q \u} & 0 & -e^{-i 2 \q \u} \\
 \kappa_1 & 1& e^{i 2 \q (\u-\v)}  & 0 & 0 & 0 & 0 & 1 & 0 & 0 & 0 & 0 \\
 0 & 1 & -1 & 0 & 0 & 0 & 0 & 0 &  1& 0 & 0 & \kappa^*_2 \\
 0 & -1 & 0 & 0 & 0 & 0 & 0 & 1 & 1 & 0 & \kappa^*_3 & 0 \\
 0 & 0 & 1 & 0 & 0 & 0 & 0 & -e^{i 2 \q (\v-\u)} & 1 &  \kappa^*_1  & 0 & 0  \\
 e^{i 2 \q \u} & 0 & 0 & 1 & 0 & 1 & -e^{i 2 \q (\u-\v)} & 0 & 0 & -1 &  e^{i 2 \q (\u-\v)} & 0 \\
 0 & 0 & 0 & 0 & 1 & 1 & 1 & 0 & 0 & 1& 1 & 1 \\
 0 & 0 & e^{i 2 \q \u} & 0 & 0 & 0 &  \kappa_1& -1& 1 & 0 & 0 & 0 \\
 0 & -e^{i 2 \q \v} & 0 & 0 & 0 &   \kappa_3 & 0 & e^{i 2 \q (\v-\u)} & 1 & 0 & 0 & 0 \\
 0 & e^{i 2 \q \v} & -e^{i 2 \q \u} & 0 & \kappa_2 & 0 & 0 & 0 & 1 & 0 & 0 & 0
\end{smallmatrix} \right]
\label{Jk}
\end{equation}
Here we have abbreviated $\kappa_1 \equiv t (1+ e^{i 2 \q \u}+e^{i 2 \q (\u-\v)})$, $\kappa_2 \equiv t (1+ e^{i 2 \q \u}+e^{i 2 \q \v})$ and $\kappa_3 \equiv t (1+ e^{i 2 \q \v}+e^{i 2 \q (\v-\u)})$, which are the additional terms coming from the next nearest neighbor interaction that is allowed by the PSG and $t$ denotes the relative strength of the interaction with respect the the nearest neighbor coupling $J=1$. 
Diagonalizing $J^{(ij)}_\q$ results in three dispersing bands, each being four-fold degenerate. Depending on the sign of the next-nearest neighbor interaction $t$, we get different transitions to valence bond solids that we are going to discuss in the following.

\subsection{Ferromagnetic n.n.n.\ interactions}
\label{sec:ferro}

For ferromagnetic next-nearest neighbor interactions  ($t>0$) the dispersion minimum of the lowest band is at zero momentum $\q=0$. The corresponding eigenvalue of the interaction matrix $J^{(ij)}_\mathbf{0}$ is given by $\lambda_+ = (3 t +\sqrt{24+9 t^2})/2$ and one choice for the four degenerate eigenvectors is
\beqn
\setcounter{MaxMatrixCols}{12}
v^{(1)} &=& \left[ \begin{matrix}
\frac{1}{\lambda_-} & 1& 0& \frac{1}{\lambda_-} & \frac{\lambda_+}{6} & \frac{-1}{\lambda_-} & 0 & 0 & 0 & 0 & \frac{-1}{\lambda_-} & \frac{\lambda_+}{6}
\end{matrix} \right]   \sqrt{6/(\lambda_+^2+6)} \notag \\
v^{(2)} &=& \left[ \begin{matrix}
\frac{1}{\lambda_-} & 0 & 1 & \frac{1}{\lambda_-} & \frac{-\lambda_+}{6} &0 & \frac{1}{\lambda_-}  & 0 & 0 & \frac{1}{\lambda_-}& 0 & \frac{-\lambda_+}{6}
\end{matrix} \right]  \sqrt{6/(\lambda_+^2+6)} \notag \\
v^{(3)} &=& \left[ \begin{matrix}
\frac{1}{\lambda_-} & 0 & 0 & \frac{1}{\lambda_-} & 0 & \frac{1}{\lambda_-}  & \frac{-1}{\lambda_-}  & 1 & 0 & \frac{-1}{\lambda_-} & \frac{1}{\lambda_-} & 0
\end{matrix} \right]  \sqrt{6/(\lambda_+^2+6)} \notag \\
v^{(4)} &=& \left[ \begin{matrix}
0 & 0 & 0 & 0 & \frac{\lambda_+}{6} & \frac{\lambda_+}{6} & \frac{\lambda_+}{6} & 0 & 1 & \frac{\lambda_+}{6} & \frac{\lambda_+}{6} & \frac{\lambda_+}{6}
\end{matrix} \right]  \sqrt{6/(\lambda_+^2+6)} \notag 
\eeqn
with $\lambda_- =  (3 t -\sqrt{24+9 t^2})/2$. This set of eigenvectors forms an orthonormal basis for the four critical modes at the transition to the confined phase. In the magnetically ordered state the magnetization $\phi_j(\R)$ at lattice site $\R=2 n \u + 2 m \v$ ($n,m \in \mathbb{N}$) and sublattice site $j \in \{1,...,12 \} $ is given by
\beq
\phi_j(\R) = \sum_{n=1...4} \psi_n v^{(n)}_j
\eeq
and is independent of the lattice site $\R$ for ferromagnetic n.n.n.\ interactions, {\em i.e.\/} the ordered VBS phases have a 12-site unit cell. The values of the four mode amplitudes $\psi_n$ are obtained by minimizing the Ginzburg-Landau functional, which in turn is given by all homogeneous polynomials in the mode amplitudes $\psi_n$ that are invariant under the PSG transformations. In order construct these polynomials we have to determine how the mode amplitudes transform under the PSG. This can be done by looking at the transformation properties of the eigenvectors $v^{(n)}$. For example, the transformation properties of the mode amplitudes with respect to translations $T_\u$ are determined via
\beq
\hat{T}_\u \phi_j = \sum_n \psi_n \, \hat{T}_\u v_j^{(n)} = \sum_{n,m}  (T_\u)_{mn}  \psi_n v_j^{(m)} = \sum_m (\hat{T}_\u \psi)_m v_j^{(m)} \ .
\eeq
The resulting PSG transformation matrices for the amplitudes of the four critical modes with respect to translations ($T_\u$), reflections ($I_x$) and rotations ($R_6$) are given by
\beq
T_\u = \begin{bmatrix}
0 & 0 & -1 & 0 \\
0 & 0 & 0 & -1 \\
1 & 0 & 0 & 0 \\
0 & 1 & 0 & 0
\end{bmatrix}, \
I_x = \begin{bmatrix}
0 & 1 & 0 & 0 \\
1 & 0 & 0 & 0 \\
0 & 0 & 1 & 0 \\
0 & 0 & 0 & -1
\end{bmatrix}, \
R_6 = \begin{bmatrix}
0 & 0 & 1 & 0 \\
1 & 0 & 0 & 0 \\
0 & 1 & 0 & 0 \\
0 & 0 & 0 & 1
\end{bmatrix} \ .
\eeq
We used the GAP program\cite{gap} to show that these three matrices generate a finite, 48 element subgroup of O(4) which is isomorphic to ${\rm GL}(2,\mathbb{Z}_3)$. We also determined that this group is isomorphic to the group $\pm \frac{1}{2} [ O \times C_2 ]$
in the classification of Conway and Smith.\cite{conway}

Next we determined the most general Ginzburg-Landau (GL) functional of the $\psi_n$.
It turns out that there are three fourth order polynomials that are invariant under this group, thus our functional for the four mode amplitudes depends on the coupling constants, $r$, $u$, $a$ and $b$, and takes the form
\beqn
\mathcal{L} &=& \sum_{n=1...4}((\nabla \psi_n)^2 + (\partial_\tau \psi_n)^2 + r \psi_n^2 + u \psi_n^4) + a \sum_{n < m} \psi_n^2 \psi_m^2  
\nonumber \\
&~& + b \big[ \psi_1^2 (\psi_2 \psi_3-\psi_2 \psi_4+\psi_3 \psi_4)  
+ \psi_2^2 (\psi_1 \psi_3+\psi_1 \psi_4-\psi_3 \psi_4) + \psi_3^2 (\psi_1 \psi_2-\psi_1 \psi_4+\psi_2 \psi_4) 
\nonumber \\
&~&~~~~~~  - \psi_4^2 (\psi_1 \psi_2+\psi_1 \psi_3+\psi_2 \psi_3) \big]. \label{GL1}
\eeqn
At $b=0$ and $a=2u$, this is just the well-known $\phi^4$ field theory with O(4) symmetry, whose critical properties are described by the 
extensively studied Wilson-Fisher fixed point. The $b$ coupling breaks the O(4) symmetry down to ${\rm GL}(2,\mathbb{Z}_3)$.
Remarkably, the renormalization group (RG) properties of just such a quartic coupling have been studied earlier by 
Toledano {\em et al.};\cite{toledano} they denoted this symmetry class as $[D_3/C_2 ; O/D_2]$, following the analysis of Du Val.\cite{duval}
Toledano {\em et al.} found that the O(4) fixed point was unstable, but were unable to find a stable critical fixed
point at two-loop order. It would be interesting to extend the analysis of Eq.~(\ref{GL1}) to higher loop order, and search for
a suitable fixed point by the methods reviewed in Ref.~\onlinecite{vicari}.

\begin{figure}
  \begin{center}
    \includegraphics[width=10cm]{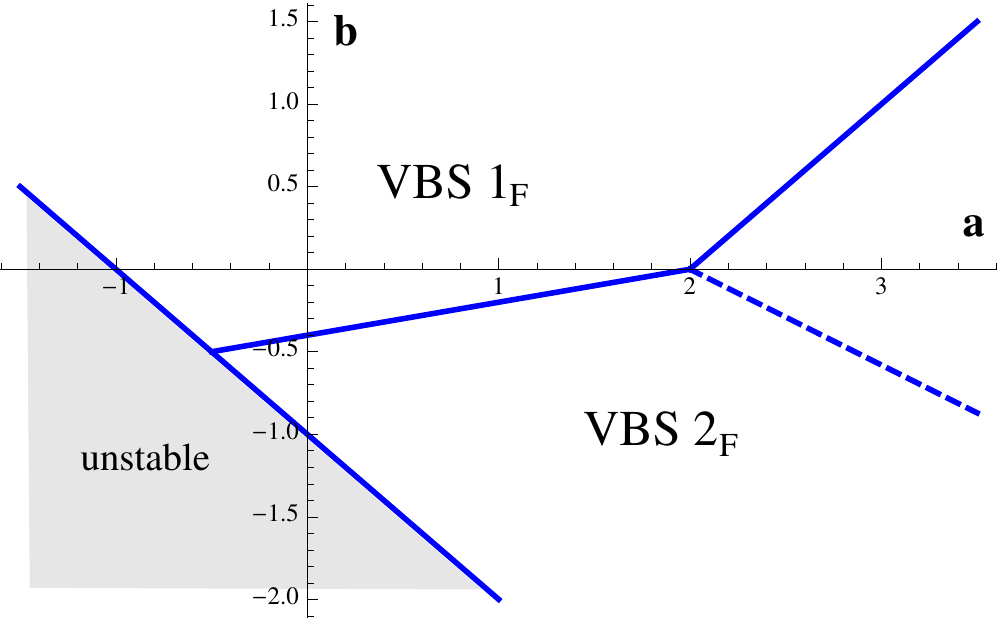}
    \caption{Phase diagram of Eqn.\ \eqref{GL1} as a function of the two couplings $a$ and $b$ (here we have set $u=1$). The VBS $1_\text{F}$ phase is not reflection symmetric, whereas the VBS $2_\text{F}$ phase is reflection symmetric. In the VBS $2_\text{F}$  phase there is a crossover (indicated by the dashed line) from a phase where one of the $\psi_n$'s is zero (left of the dashed line) to a phase where three of the $\psi_n$'s are zero (right of the dashed line).}
    \label{fig:PD1}
  \end{center}
\end{figure}

Despite the difficulty in finding a suitable critical fixed point, we can assume the existence of a second-order quantum phase transition,
and use Eq.~(\ref{GL1}) to determine the structures of possible confining phases.
Minimizing this functional for the magnetically ordered phase $r<0$ gives rise to two possible phases, depending on the values of the two parameters $a$ and $b$ in the GL-functional \eqref{GL1}. The phase diagram is shown in Fig.\ \ref{fig:PD1}. 

\begin{figure}
\begin{center}
\includegraphics[width=10cm]{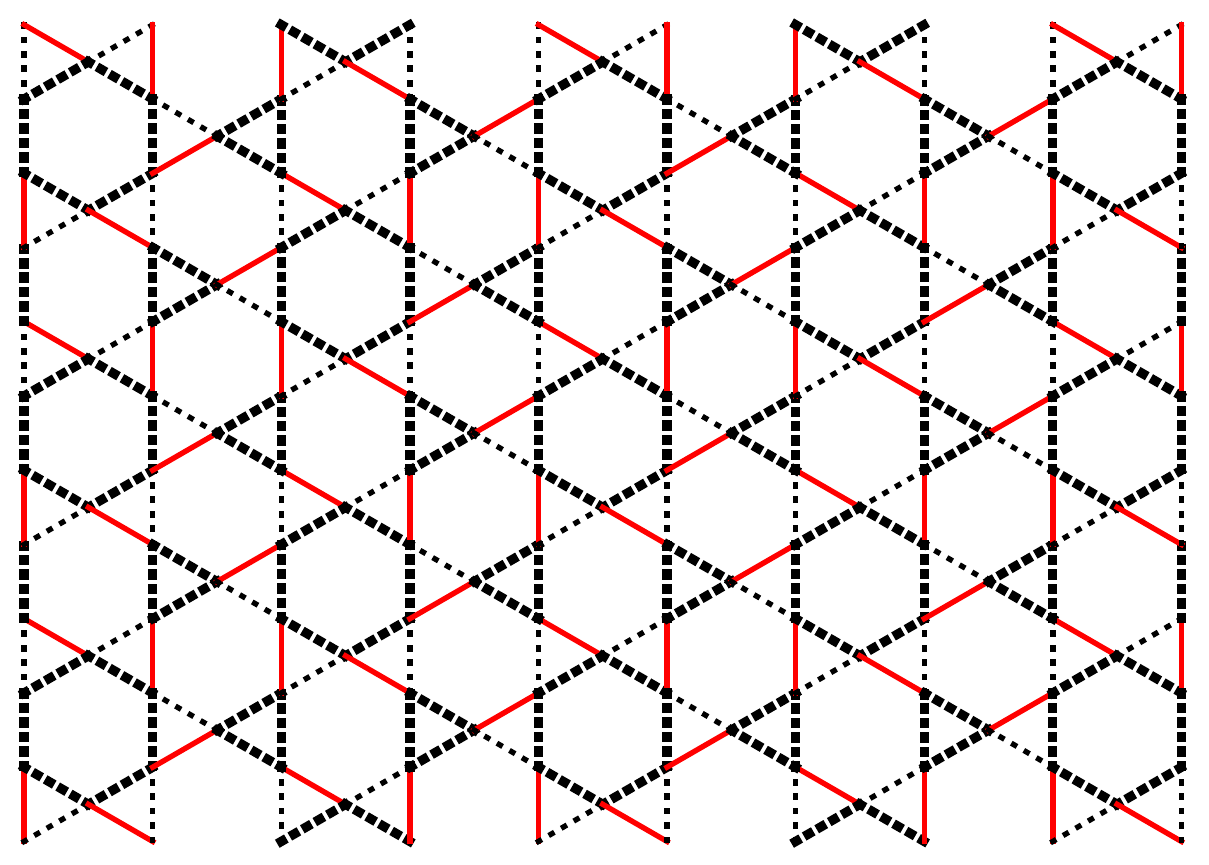}
\end{center}
\caption{Bond pattern in the VBS $1_\text{F}$  phase. Plotted is the gauge invariant bond-strength $J_{ij} \phi_i \phi_j$ for nearest neighbor bonds on the dice-lattice at $a=b=1$, which has been assigned to each respective kagome bond. Black dashed lines indicate satisfied bonds ($-J_{ij} \phi_i \phi_j < 0$), red solid lines are frustrated bonds  ($-J_{ij} \phi_i \phi_j > 0$). The thickness of the bonds is proportional to the bond-strength.}
\label{figvbs1}
\end{figure}

In the VBS $1_\text{F}$ phase the GL-functional \eqref{GL1} has 16 degenerate minima corresponding to different magnetization patterns. Note, however, that the magnetization itself is not a gauge invariant quantity. In fact, all degenerate minima give rise to the same bond patterns, which are related by simple lattice symmetry transformations. The bond pattern in the VBS $1_\text{F}$  phase is symmetric under rotations by $\pi/3$, translations by $2\u$ and $2\v$ but it is not reflection symmetric. The 16 minima thus correspond to 8 different bond patterns that are related by translations by $\u$ and $\v$ as well as by a reflection about the x-axis (the factor of two arises from a global spin flip symmetry). In Fig.\ \ref{figvbs1} we show the bond pattern in the VBS $1_\text{F}$  phase, {\em i.e.\/} we plot the gauge-invariant value of the bond-strength $J_{ij} \phi_i \phi_j$ for every nearest neighbor bond. In this figure we mapped the bond pattern from the dice lattice back to the kagome lattice by assigning the value of the bond-strength on the dice lattice to the bond on the kagome lattice that intersects the dice lattice bond. Deep in the confined phase the bonds plotted in Fig.\ \ref{figvbs1} thus represent the expectation value $\langle \sigma^x_{ij} \rangle$ of the $\mathbb{Z}_2$ gauge field on the kagome bonds in the transverse field direction. 

One way to relate the symmetries of valence bond solid phases on the kagome lattice to the ordered phases of the frustrated Ising model is to make use of the mapping between hardcore dimer models and frustrated Ising models \cite{moessner}, {\em i.e.\/} by putting a dimer on every bond of the kagome lattice that intersects a frustrated bond on the dual dice lattice. The dimer covering that is obtained in this way for the VBS $1_\text{F}$  phase is shown in Fig.\ \ref{figkagome}. In comparison to previously obtained dimer coverings of the kagome lattice\cite{nikolic, didier3} which maximize the number of perfectly flippable hexagons on a 36-site unit cell, this dimer covering maximizes the number of perfect flippable diamonds on a 12-site unit cell. Note that this bond pattern is similar to the diamond pattern observed by Yan {\em et al.} \cite{white}. 
It is important to note, that this mapping between the frustrated Ising model \eqref{HIsing} and the corresponding hardcore dimer model is strictly valid only in the limit where the transverse field vanishes, \emph{i.e.\ } deep in the ordered phase. In this regime our GL-approach is quantitatively not reliable, however. It is thus no surprise, that not all of our dimer coverings are hardcore coverings (see \emph{e.g.\ } the right dimer covering shown in Fig.\ \ref{figkagome}). The GL-approach works well to determine broken symmetries of VBS states as well as the critical properties close to the confinement transition, where it is sufficient to keep only the relevant terms up to fourth order in the functional. On general grounds it would be necessary to include all higher order terms in the GL-functional in order to obtain reliable hardcore dimer coverings in the limit of small transverse fields. Apparently, the VBS $1_\text{F}$ phase is an exception to this rule.

We note that the VBS $1_\text{F}$ state has recently been identified as the ground state of the deformed kagome lattice spin-$1/2$ antiferromagnet Rb$_2$Cu$_3$SnF$_{12}$.\cite{matan} It is possible that the spin physics being discussed here played a role in the lattice
distortion observed in this experiment.\cite{ybkim} Also in Zn-paratacamite, Zn$_x$Cu$_{4-x}$(OH)$_6$Cl$_2$, there is a 
transition\cite{shlee} between
a spin-liquid phase near $x=1$ to a distorted kagome lattice near $x=0$, and it has been argued\cite{lawler} 
that a `pinwheel' VBS state, which is identical
in symmetry to our  VBS $1_\text{F}$ state, plays a role in the latter case.

\begin{figure}
\begin{center}
\includegraphics[width=6.5cm]{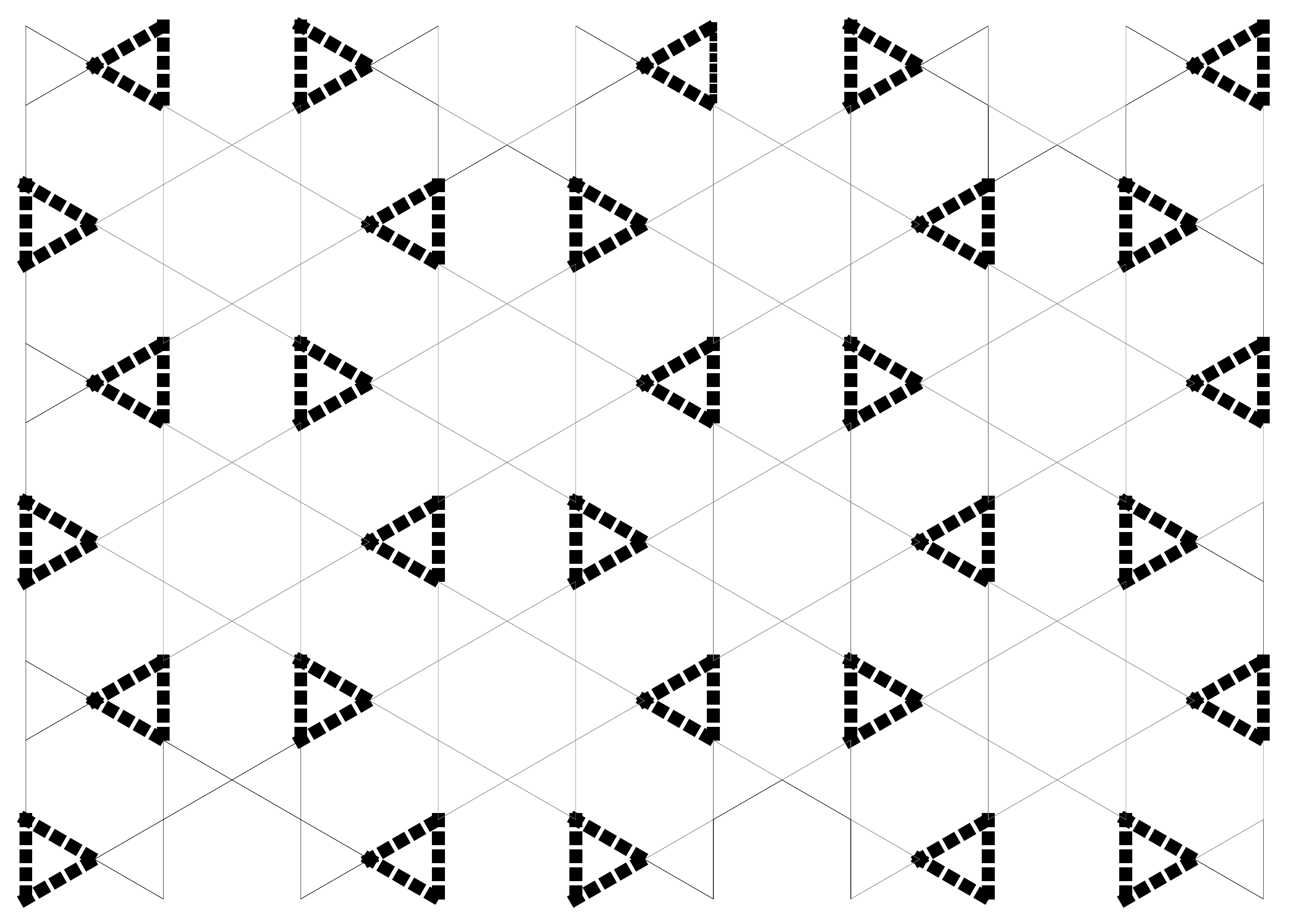}
\hspace{1cm}
\includegraphics[width=6.5cm]{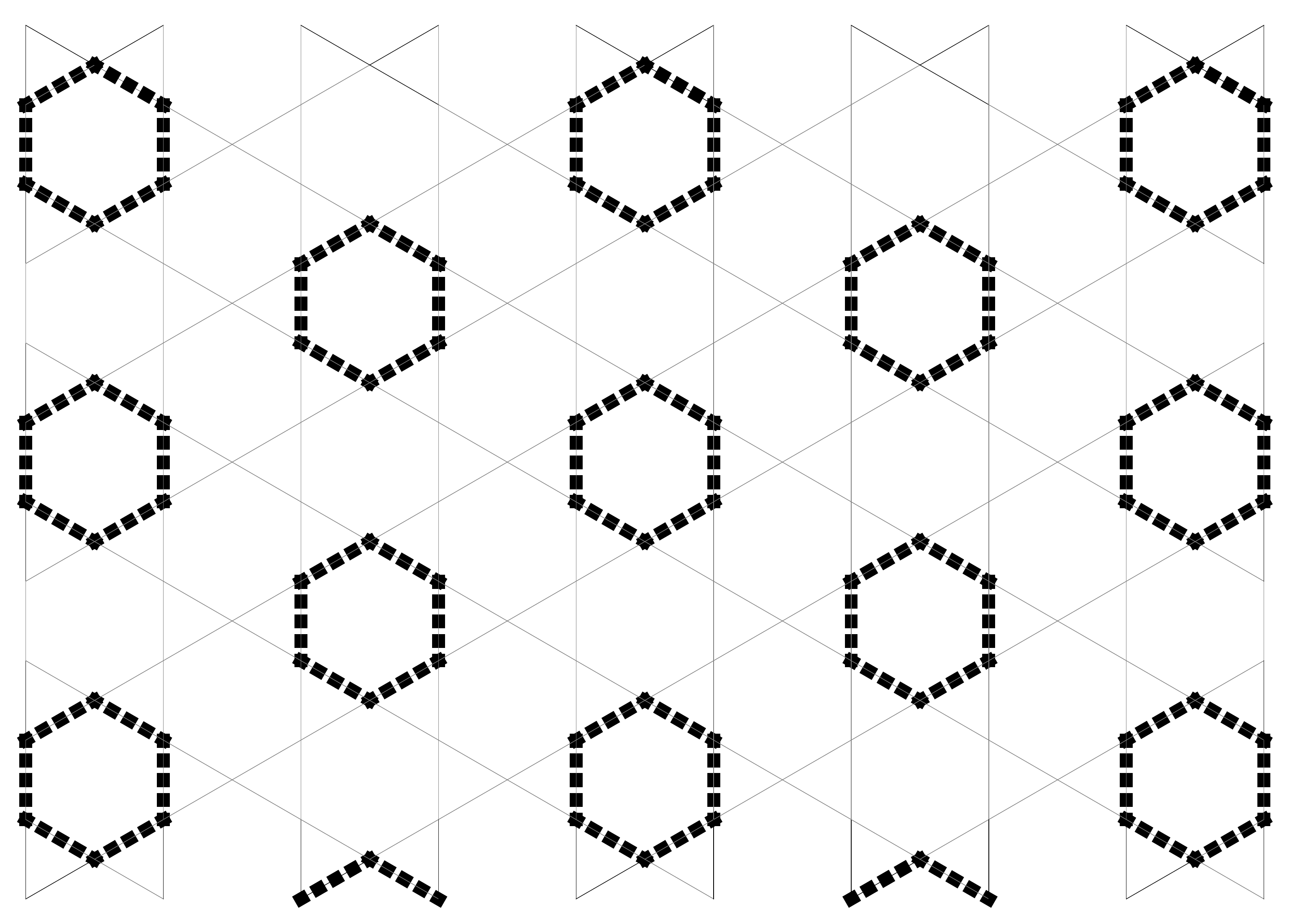}
\end{center}
\caption{Bond patterns in the VBS $2_\text{F}$  phase. Plotted is the gauge invariant bond-strength $J_{ij} \phi_i \phi_j$ for nearest neighbor bonds on the dice lattice, shown on the corresponding kagome bonds. Black dashed lines represent satisfied bonds ($-J_{ij} \phi_i \phi_j < 0 $). In this phase there are no frustrated bonds at all. The thickness of the bonds is proportional to the bond-strength.}
\label{figvbs2}
\end{figure}

In the VBS $2_\text{F}$  phase the GL-functional \eqref{GL1} has 8 degenerate minima and the corresponding bond-patterns have an additional reflection symmetry as compared to the VBS $1_\text{F}$  phase. Moreover, there is a crossover from a phase where one of the $\psi_n$'s is zero to a phase where three of the $\psi_n$'s are zero, indicated by the dashed line in Fig.\ \ref{fig:PD1}. The corresponding bond patterns are shown in Fig.\ \ref{figvbs2}.

\subsection{Antiferromagnetic n.n.n.\ interactions}
\label{sec:antiferro}

If the additional next-nearest neighbor interactions are antiferromagnetic ($t<0$) and smaller than a critical value $|t|< t_c \sim 1$, the dispersion-minimum of the lowest band lies at the edges of the Brillouin zone, {\em i.e.\/} at $\q = \pm \mathbf{Q}_1 =  \big( 0, \ \pm \frac{2 \pi}{3 \sqrt{3}} \big)$ for our gauge choice with a  hexagonal 12-site unit cell. In this case eight modes become critical at the confinement transition and the resulting unit cell has 36 sites. The corresponding eigenvalue of the interaction matrix $J^{(ij)}_{\pm \mathbf{Q}_1}$ is $\sqrt{6}$ and the eight eigenvectors occur in complex conjugate pairs
\beqn
\setcounter{MaxMatrixCols}{12}
v^{(1)}_{Q_1} = v^{(1)*}_{-Q_1} &=& \left[ \begin{matrix}
\frac{1}{\sqrt{12}} & \frac{1}{\sqrt{2}}& 0& \frac{1}{\sqrt{12}} & \frac{1}{\sqrt{12}} & \frac{-1}{\sqrt{12}} & 0 & 0 & 0 & 0 & \frac{e^{i \pi/3}}{\sqrt{12}} & \frac{-e^{i \pi/3}}{\sqrt{12}}
\end{matrix} \right]  \notag \\
v^{(2)}_{Q_1} = v^{(2)*}_{-Q_1} &=& \left[ \begin{matrix}
\frac{1}{\sqrt{12}} & 0 & \frac{1}{\sqrt{2}} & \frac{-e^{i \pi/3}}{\sqrt{12}} & \frac{-1}{\sqrt{12}} & 0 & \frac{1}{\sqrt{12}} & 0 & 0 & \frac{-e^{-i \pi/3}}{\sqrt{12}} & 0 & \frac{e^{-i \pi/3}}{\sqrt{12}}
\end{matrix} \right]  \notag \\
v^{(3)}_{Q_1} = v^{(3)*}_{-Q_1}&=& \left[ \begin{matrix}
\frac{-e^{i \pi/3}}{\sqrt{12}} & 0 & 0& \frac{1}{\sqrt{12}} & 0 & \frac{1}{\sqrt{12}} & \frac{e^{-i \pi/3}}{\sqrt{12}} & \frac{1}{\sqrt{2}} & 0 & -\frac{1}{\sqrt{12}} & \frac{-e^{-i \pi/3}}{\sqrt{12}} & 0
\end{matrix} \right]  \notag \\
v^{(4)}_{Q_1} = v^{(4)*}_{-Q_1}&=& \left[ \begin{matrix}
0 & 0 & 0& 0 & \frac{1}{\sqrt{12}} & \frac{1}{\sqrt{12}} & \frac{1}{\sqrt{12}} & 0 & \frac{1}{\sqrt{2}} & \frac{1}{\sqrt{12}} & \frac{1}{\sqrt{12}} & \frac{1}{\sqrt{12}}
\end{matrix} \right]  \notag 
\eeqn
Note that in this case the magnetization at lattice site $\R=2 n \u +2 m \v$ and sublattice site $j$ is given by
\beq
\phi_j(\R) =  e^{i \Q_1 \cdot \R} \sum_{n=1...4} \psi_n v^{(n)}_{\Q_1,  j} + \text{c.c.}
\label{magnetiz}
\eeq
The PSG transformations of the four complex mode amplitudes $\psi_n$ can be determined similarly to the ferromagnetic case. Quite generally they are defined by 
\beq
\hat{\mathcal{O}} \phi_j(\R) = \text{Re} \Big[ e^{i \Q_1 \cdot (\hat{\mathcal{O}} \R)} \sum_{n=1...4} \psi_n \, \hat{\mathcal{O}} v^{(n)}_{\Q_1,  j} \Big] \doteq  \text{Re} \Big[ e^{i \Q_1 \cdot \R} \sum_{n=1...4} (\hat{\mathcal{O}} \psi)_n \, v^{(n)}_{\Q_1,  j} \Big] 
\eeq
If we define the vector $\Psi=\big(\psi_1,..,\psi_4,\psi_1^*,..,\psi_4^* \big)$ , the PSG transformation matrices corresponding to translations by $\u$, reflections about the x-axis and rotations by $\pi/3$ around the central site, that act on the vector $\Psi$ take the form 
\beq
T_\u = \begin{bmatrix}
0 & 0 & -1 & 0 & 0 & 0 & 0 & 0\\
0 & 0 & 0 & -1 & 0 & 0 & 0 & 0\\
e^{i 2 \pi/3} & 0 & 0 & 0 & 0 & 0 & 0 & 0 \\
0 & e^{i 2 \pi/3} & 0 & 0 & 0 & 0 & 0 & 0 \\
0 & 0 & 0 & 0 & 0 & 0 & -1 & 0 \\
0 & 0 & 0 & 0 & 0 & 0 & 0 & -1 \\
0 & 0 & 0 & 0 & e^{-i 2 \pi/3}  & 0 & 0 & 0 \\
0 & 0 & 0 & 0 & 0 & e^{-i 2 \pi/3}  & 0 & 0 \\
 \end{bmatrix},
\eeq
\beq
I_x = \begin{bmatrix}
0 & 0 & 0 & 0 & 0 & 1 & 0 & 0\\
0 & 0 & 0 & 0 & 1 & 0 & 0 & 0\\
0 & 0 & 0 & 0 & 0 & 0 & e^{-i 2 \pi/3} & 0 \\
0 & 0 & 0 & 0 & 0 & 0 & 0 & -1 \\
0 & 1 & 0 & 0 & 0 & 0 & 0 & 0 \\
1 & 0 & 0 & 0 & 0 & 0 & 0 & 0 \\
0 & 0 & e^{i 2 \pi/3} & 0 &0  & 0 & 0 & 0 \\
0 & 0 & 0 & -1 & 0 & 0  & 0 & 0 \\
 \end{bmatrix},
\eeq
\beq
R_6 = \begin{bmatrix}
0 & 0 & 0 & 0 & 0 & 1 & 0 & 0\\
0 & 0 & 0 & 0 & 0 & 0 &e^{-i 2 \pi/3} & 0\\
0 & 0 & 0 & 0 & 1 & 0 &0 & 0 \\
0 & 0 & 0 & 0 & 0 & 0 & 0 & 1 \\
0 & 1 & 0 & 0 & 0 & 0 & 0 & 0 \\
0 & 0 & e^{i 2 \pi/3} & 0 & 0 & 0 & 0 & 0 \\
1 & 0 & 0 & 0 &0  & 0 & 0 & 0 \\
0 & 0 & 0 & 1 & 0 & 0  & 0 & 0 \\
 \end{bmatrix}.
\eeq
As in the previous subsection, we used the GAP program to determine that 
these three matrices generate a 288 element subgroup of O(8) which is isomorphic to ${\rm GL}(2,\mathbb{Z}_3) \times {\rm D}_3$
The Ginzburg-Landau functional is again given by all homogeneous polynomials that are invariant under this group. At fourth order there are five such polynomials, thus the GL-functional depends on the coupling constants $r, u, a_1,...,a_4$ and is given by
\beqn
\mathcal{L}_4 &=&\sum_{n=1..4} \big(r  \psi_n^2 + u \psi_n^4 \big)  +\psi_2^2 \psi_3^2  \Big[ a_1-2 a_4 \cos(2 (\theta_2-\theta_3)) \Big] \notag \\
&&+\psi_1 \psi_2^2 \psi_4 \Big[-a_2 \cos(\theta_1-\theta_4)-2 a_3 \cos(\theta_1-2 \theta_2+\theta_4) - \sqrt{3} a_2 \sin(\theta_1-\theta_4) \Big] \notag \\
&&+\psi_1 \psi_3^2 \psi_4 \Big[a_2
   \cos(\theta_1-\theta_4)+2 a_3 \cos(\theta_1-2 \theta_3+\theta_4)+\sqrt{3} a_2 \sin(\theta_1-\theta_4)\Big] \notag \\
 && +\psi_1^2 \psi_4^2 \Big[a_1+a_4 \cos(2 (\theta_1-\theta_4))  -\sqrt{3} a_4 \sin(2 (\theta_1-\theta_4))\Big] \notag \\
&& + \psi_1 \psi_2 \psi_4^2 \Big[a_2 \cos(\theta_1-\theta_2)+2 a_3 \cos(\theta_1+\theta_2-2 \theta_4)-\sqrt{3} a_2 \sin(\theta_1-\theta_2)\Big] \notag \\
&& + \psi_1^2 \psi_2^2 \Big[a_1+a_4 \cos(2 (\theta_1-\theta_2))+\sqrt{3} a_4 \sin(2 (\theta_1-\theta_2))\Big] \notag \\ 
&& +\psi_1^2 \psi_2
   \psi_4 \Big[2 a_3 \cos(2 \theta_1-\theta_2-\theta_4)+a_2 \cos(\theta_2-\theta_4)-\sqrt{3}
   a_2 \sin(\theta_2-\theta_4)\Big] \notag \\
 && + \psi_2^2 \psi_4^2 \Big[a_1+a_4 \cos(2 (\theta_2-\theta_4))+\sqrt{3} a_4 \sin(2 (\theta_2-\theta_4))\Big] \notag \\
 && +\psi_2 \psi_3^2 \psi_4 \Big[-a_2
   \cos(\theta_2-\theta_4)+a_3 \cos(\theta_2-2 \theta_3+\theta_4) +\sqrt{3} a_2 \sin(\theta_2-\theta_4) \notag \\
   && \ \ \ \ \ +\sqrt{3} a_3 \sin(\theta_2-2 \theta_3+\theta_4)\Big] \notag \\
   && +\psi_1 \psi_2 \psi_3^2
   \Big[-a_2 \cos(\theta_1-\theta_2)+a_3 \cos(\theta_1+\theta_2-2
   \theta_3)+\sqrt{3} a_2 \sin(\theta_1-\theta_2) \notag \\
   && \ \ \ \ \ -\sqrt{3} a_3 \sin(\theta_1+\theta_2-2 \theta_3)\Big] \notag \\
   && +\psi_1 \psi_3 \psi_4^2 \Big[a_2
   \cos(\theta_1-\theta_3)+2 a_3 \cos(\theta_1+\theta_3-2 \theta_4)-\sqrt{3} a_2
   \sin(\theta_1-\theta_3)\Big] \notag \\
   &&+\psi_1^2 \psi_3^2 \Big[a_1+a_4 \cos(2
   (\theta_1-\theta_3))+\sqrt{3} a_4 \sin(2 (\theta_1-\theta_3))\Big] \notag \\
   &&+\psi_1^2
   \psi_2 \psi_3 \Big[a_3 \cos(2 \theta_1-\theta_2-\theta_3)+2 a_2
   \cos(\theta_2-\theta_3)+\sqrt{3} a_3 \sin(2 \theta_1-\theta_2-\theta_3)\Big] \notag \\
   &&+\psi_2^2 \psi_3 \psi_4 \Big[-a_3 \cos(2 \theta_2-\theta_3-\theta_4)+a_2 \cos(\theta_3-\theta_4)+\sqrt{3} a_3 \sin(2 \theta_2-\theta_3-\theta_4) \notag \\
   && \ \ \ \ \ -\sqrt{3} a_2 \sin(\theta_3-\theta_4)\Big] \notag \\
   &&+\psi_1^2 \psi_3 \psi_4 \Big[-2 a_3 \cos(2 \theta_1-\theta_3-\theta_4)-a_2 \cos(\theta_3-\theta_4)+\sqrt{3} a_2 \sin(\theta_3-\theta_4)\Big] \notag \\
   && +\psi_3^2
   \psi_4^2 \Big[a_1+a_4 \cos(2 (\theta_3-\theta_4))+\sqrt{3} a_4 \sin(2 (\theta_3-\theta_4))\Big] \notag \\
   &&+\psi_1 \psi_2^2 \psi_3 \Big[-a_2 \cos(\theta_1-\theta_3)+a_3
   \cos(\theta_1-2 \theta_2+\theta_3)+\sqrt{3} a_2 \sin(\theta_1-\theta_3) \notag \\
   && \ \ \ \ \ -\sqrt{3} a_3 \sin(\theta_1-2 \theta_2+\theta_3)\Big] \notag \\
   && +\psi_2 \psi_3
   \psi_4^2 \Big[-2 a_2 \cos(\theta_2-\theta_3)-a_3 \cos(\theta_2+\theta_3)-\sqrt{3} a_3 \sin(\theta_2+\theta_3)\Big] \ .
\label{GL2}
\eeqn
Here we have expressed the complex mode amplitudes in terms of their absolute value and phase $\psi_n e^{i \theta_n}$. Note that this fourth order Ginzburg-Landau functional has a remaining continuous $U(1)$ symmetry, since it is invariant under a change of all phases $\theta_n \to \theta_n + \chi$. The ``magnetization'' \eqref{magnetiz} is not invariant under this $U(1)$ transformation, however. In order to break this continuous degeneracy we need to include higher order terms in the Ginzburg-Landau functional. Among the invariant sixth order polynomials there are five which break the $U(1)$ symmetry. A full analysis of the invariant GL-functional at sixth order is beyond the scope of this paper, and we restrict ourselves to the simplest $U(1)$-breaking sixth order term. The $U(1)$-invariant sixth order terms are not expected to qualitatively change our results at the confinement transition.

\begin{figure}
  \begin{center}
    \includegraphics[width=10cm]{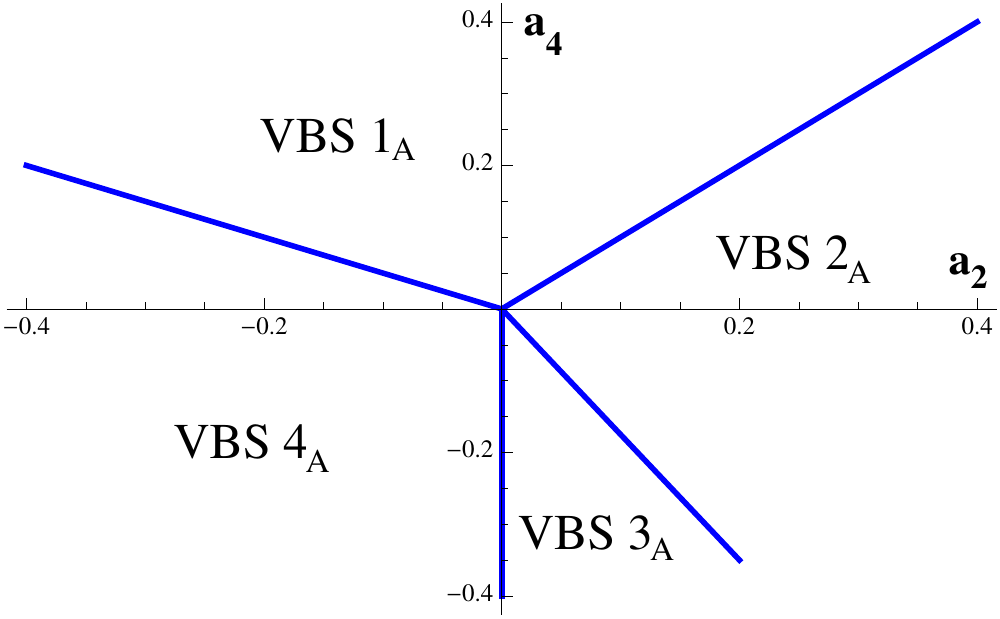}
    \caption{Phase diagram obtained from the GL-functional \eqref{GL3} as a function of the couplings $a_2$ and $a_4$. The other parameters are fixed at $u=1$, $a_1=1/2$, $a_3=a_2$, $a_5 = 1/20$ and $a_6 = -1/25$. The different phases are described in the text.}
    \label{fig:PD2}
  \end{center}
\end{figure}

\begin{figure}
\begin{center}
\includegraphics[width=6.5cm]{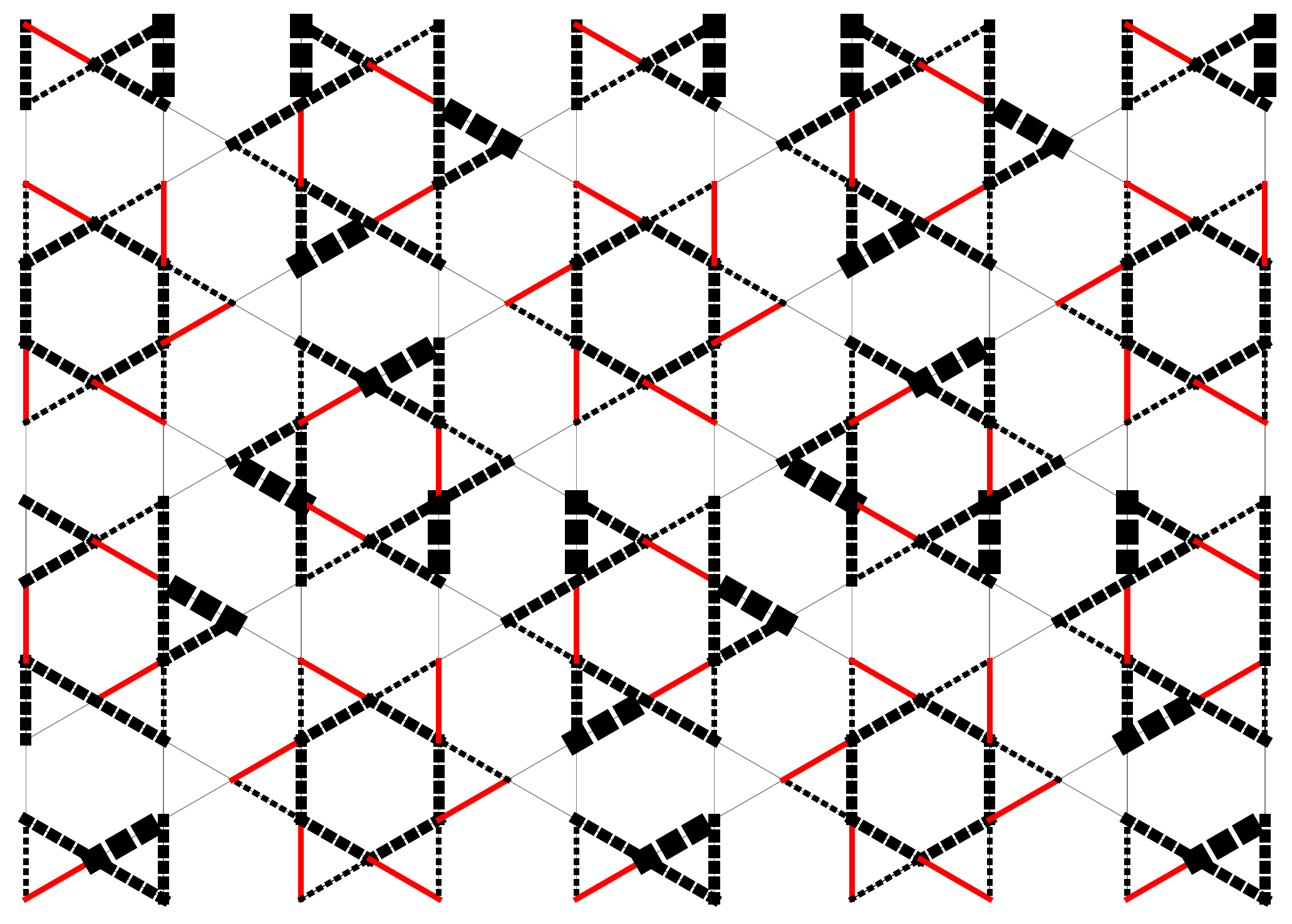}
\hspace{1cm}
\includegraphics[width=6.5cm]{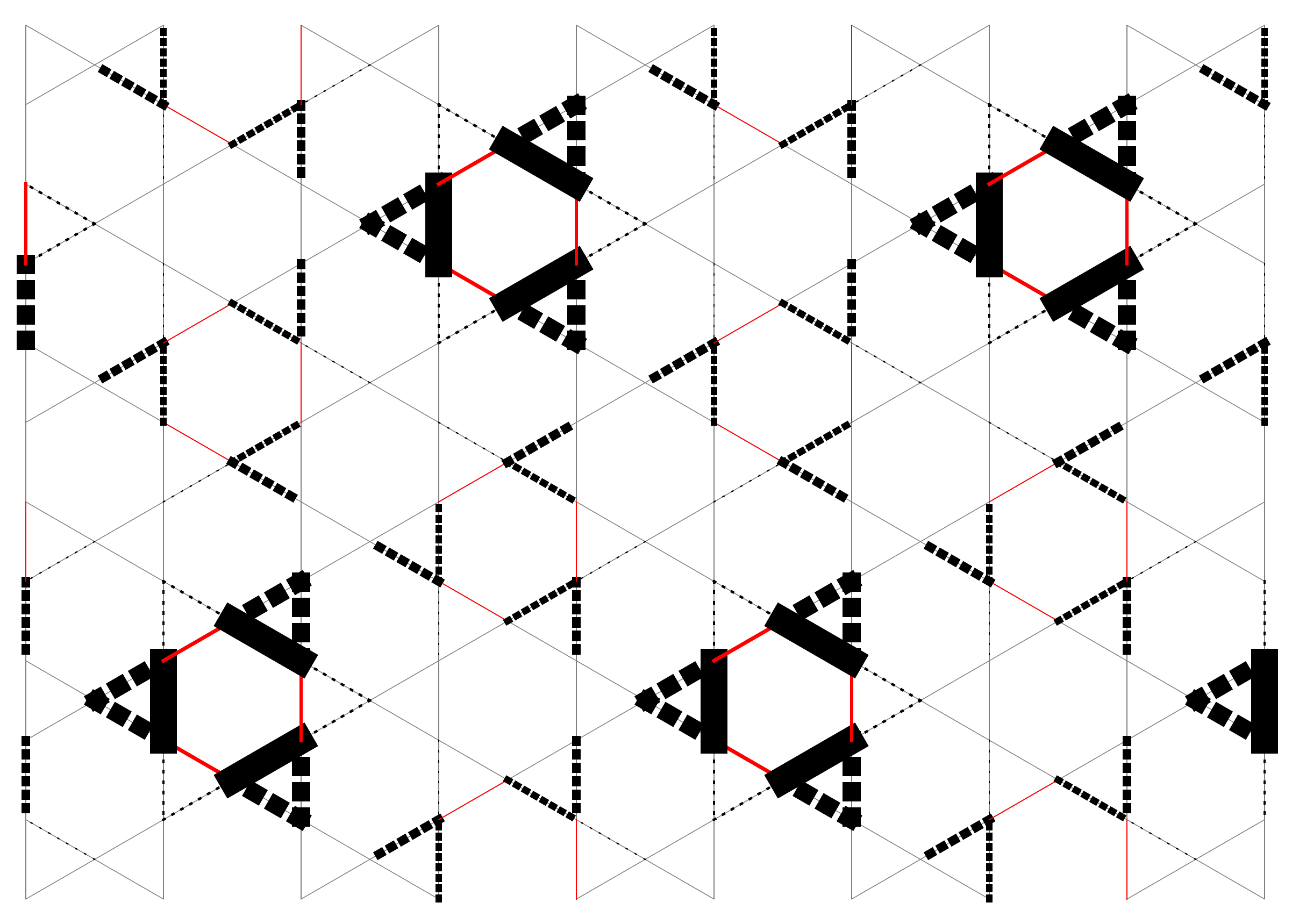}

\vspace{1cm}
\includegraphics[width=6.5cm]{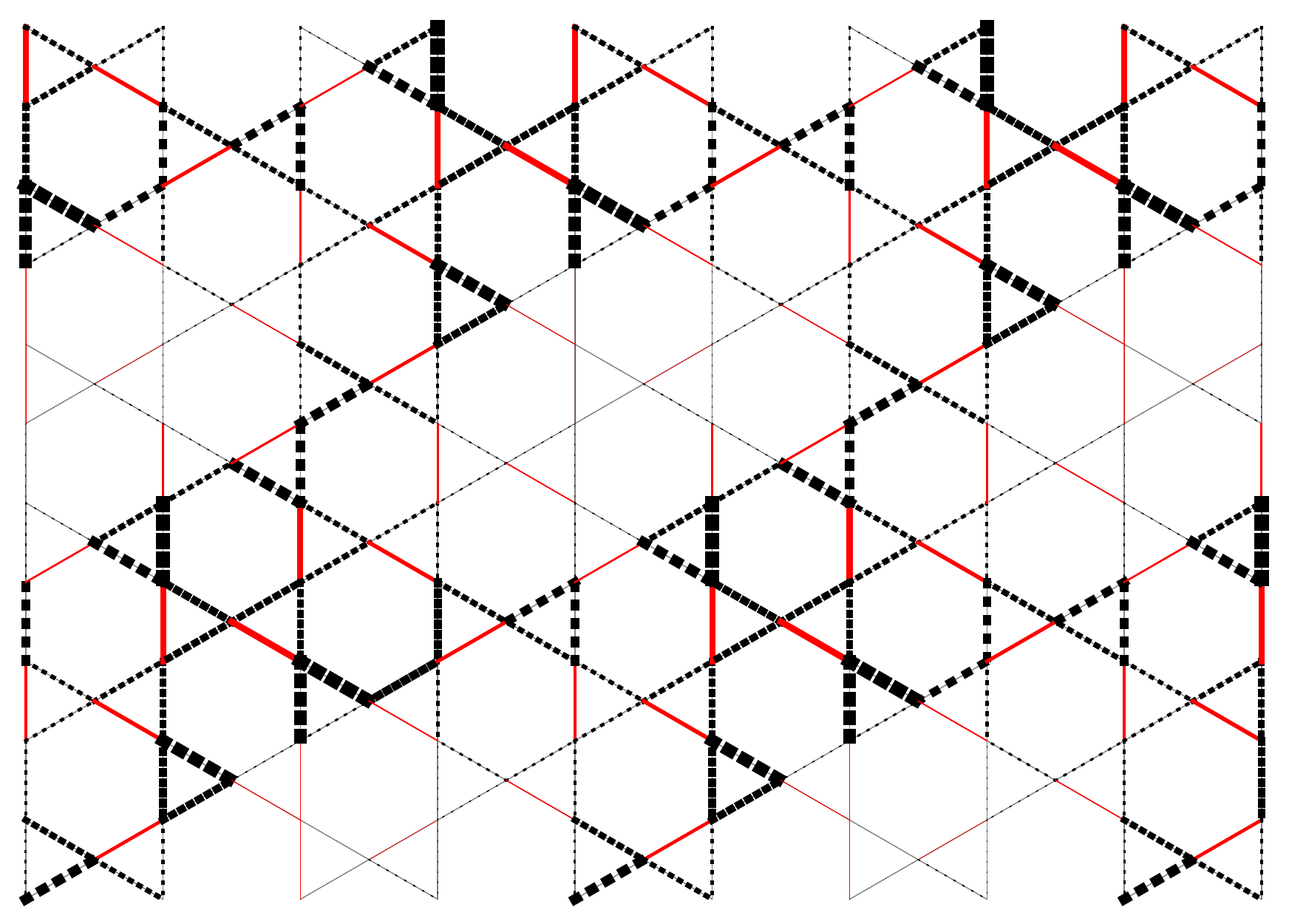}
\hspace{1cm}
\includegraphics[width=6.5cm]{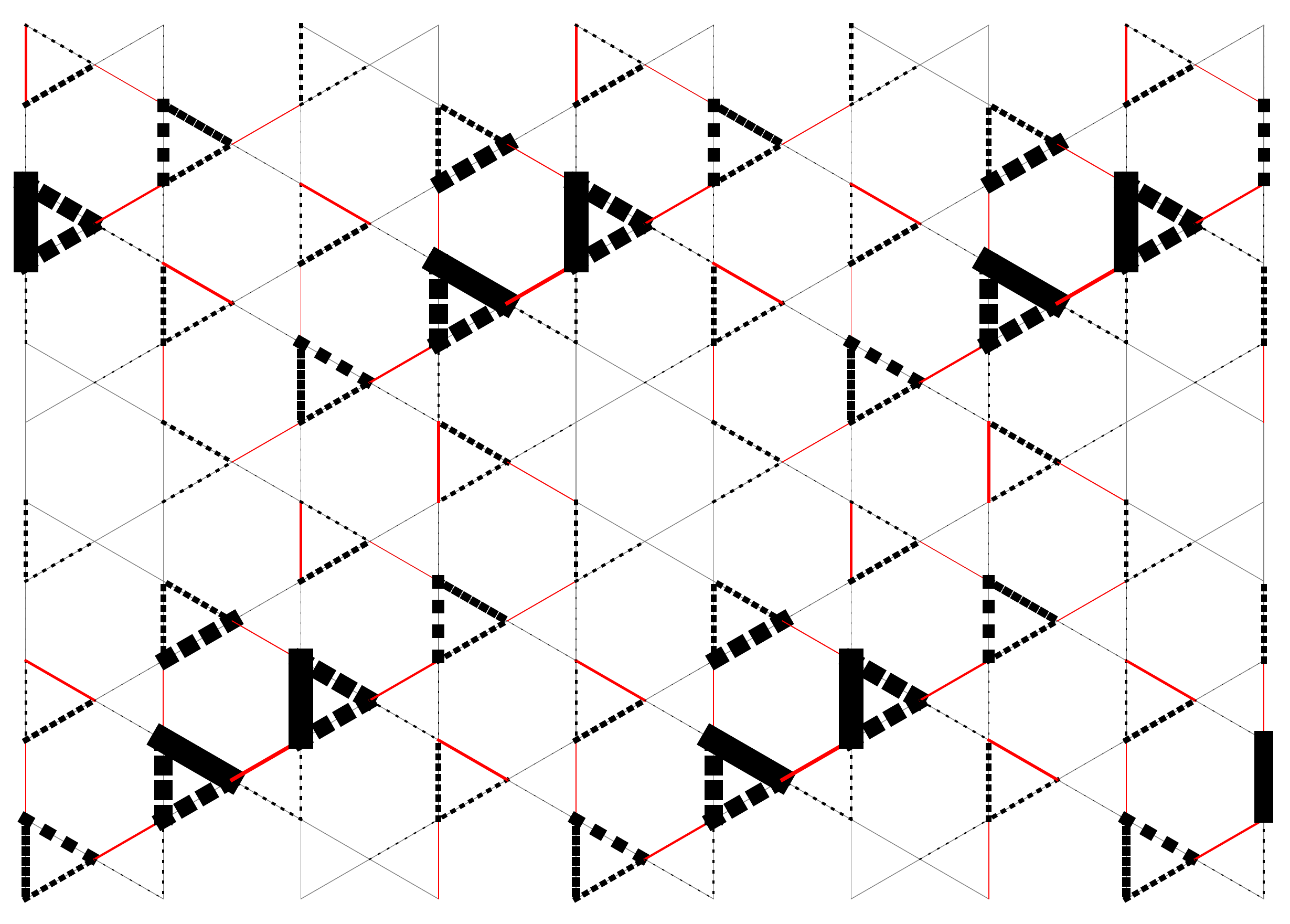}
\end{center}
\caption{Bond patterns in the VBS $1_\text{A}$,  $2_\text{A}$, $3_\text{A}$ and $4_\text{A}$ phase (clockwise, starting from the upper left). Again we plot the gauge invariant bond-strength $J_{ij} \phi_i \phi_j$ for nearest neighbor bonds on the dice lattice, mapped to the corresponding kagome bonds. Black dashed lines represent satisfied bonds ($-J_{ij} \phi_i \phi_j < 0 $), red lines are frustrated bonds. The thickness of the lines is proportional to the bond strength. }
\label{figvbsAF1}
\end{figure}

In the remainder of this section we are going to consider the following fourth order GL-functional, including the simplest invariant sixth order $U(1)$-breaking polynomial, which takes the form
\beq
\mathcal{L} = \mathcal{L}_4 + \sum_{n=1...4} \psi_n^6 \big(a_5 + a_6 \cos[6 \theta_n] \big) \ .
\label{GL3}
\eeq
In total our simplified GL-functional thus has seven coupling constants. Again, a complete analysis of the phase diagram as a function of these seven couplings is hardly feasible. A representative slice of the phase diagram is shown in Fig.\ \ref{fig:PD2}, where we have fixed the values\cite{footnote1} $u=1$, $a_1=1/2$, $a_3=a_2$, $a_5=1/20$ and $a_6=-1/25$ and show the different phases as function of the two remaining parameters $a_2$ and $a_4$. All phases have a 36-site unit cell and are invariant with respect to translations by $4\u-2\v$ and $4 \v-2 \u$. In the above mentioned parameter regime there are four different ordered phases. VBS $1_\text{A}$  has a $\pi/3$ rotational symmetry, but is not reflection symmetric. A dimer representation of this state is shown in Fig.\ \ref{figkagome}, which was obtained by putting a dimer on every kagome bond that intersects a frustrated bond on the dice lattice. This dimer covering suggests that our VBS $1_\text{A}$ state is identical to previously found valence bond solid states on the kagome lattice which maximize the number of perfectly flippable hexagons \cite{nikolic, didier3}.
The VBS $2_\text{A}$  phase is symmetric under $2 \pi/3$ rotations and has a reflection symmetry. No rotational symmetry is present in the VBS $3_\text{A}$ and $4_\text{A}$  phases. The $3_\text{A}$  phase has a reflection symmetry, however, which is not present in the $4_\text{A}$  phase. Bond patterns of all four phases are shown in Fig.\  \ref{figvbsAF1}. 

We performed a one-loop RG calculation for \eqref{GL2} and found six additional fixed points besides the Gaussian one, but all of them turned out to be unstable. Again it would be useful to revisit this issue using higher loop methods.\cite{vicari}

\section{Conclusions}
\label{sec:conc}
We have studied confinement transitions of $\mathbb{Z}_2$ spin liquids of Heisenberg antiferromagnets 
on the kagome lattice by constructing field theories that are consistent with the projective symmetry group of the vison excitations. Depending on the sign of the next-nearest neighbor interaction between the visons,
we found that the visons transformed under the group ${\rm GL}(2, \mathbb{Z}_3)$ for the simpler case, and under
${\rm GL}(2, \mathbb{Z}_3) \times {\rm D}_3$ for the other case.
Our analysis shows that  possible VBS phases close to the confinement transition are strongly constrained by the vison PSG. We found VBS states that break the translational symmetry of the kagome lattice with either 12- or 36-site unit cells, for the two vison PSGs respectively. The two possible VBS states with 12-site unit cells do not break the rotation symmetry of the kagome lattice but one of them breaks the reflection symmetry;
this state is closely connected to the ``diamond pattern'' enhancement observed in the recent numerical study
of Yan {\em et al.}\cite{white}
As far as possible VBS states with 36-site unit cells are concerned, our analysis is not exhaustive. Nevertheless, we found different VBS states with either full, reduced or no rotation symmetry, as well as states that do or do not break the reflection symmetry of the kagome lattice.
Our results should be useful in more completely characterizing spin liquids in numerical or experimental studies of the 
kagome antiferromagnet.

Analogous analyses for $Z_2$ spin liquids on other lattices have been carried out in other cases (see Appendix~\ref{app:geom}).
In all other cases, the effective theory for confining transition has an emergent continuous symmetry, and the criticality can be computed
using properties of the Wilson-Fisher fixed point; reduction to the discrete lattice symmetry appears only upon including higher-order
couplings which are formally ``irrelevant'' at the critical fixed point. The kagome lattice is therefore the unique case (so far) in which
the reduction to discrete lattice symmetry appears already in the critical theory: these is the theory in Eqs.~(\ref{GL1}), 
and its relevant quartic couplings are invariant only under discrete symmetries. This suggests
that the numerical studies of confinement transitions may be easier on the kagome lattice.

\acknowledgements

We thank Steve White for sharing the results of Ref.~\onlinecite{white} prior to publication, and for valuable discussions. 
We are also grateful to M.~Lawler and C.~Xu for discussions. This
research was supported by the National Science Foundation under
grant DMR-0757145 and by a MURI grant
from AFOSR. MP is supported by the Erwin-Schr\"odinger-Fellowship J 3077-N16 of the Austrian Science Fund (FWF). 

\appendix

\section{Berry phase of a vison}
\label{app:berry}

This appendix will compute the Berry phase of a vison moving around a $S=1/2$ spin of the antiferromagnet, as illustrated in Fig.~\ref{visonberry}.
We will follow the method of Section III.A of Ref.~\onlinecite{rsb}, generalized to a $\mathbb{Z}_2$ spin liquid as in Ref.~\onlinecite{sskag}.

We consider the time-dependent Schwinger boson Hamiltonian
\beq
\mathcal{H}_b^v (\tau)  =  - \sum_{i < j} Q_{ij}^v (\tau)  \varepsilon_{\alpha\beta} b_{i \alpha}^\dagger b_{j \beta}^\dagger + \mbox{H.c.} 
+  \sum_i \lambda_i^v (\tau) b_{i \alpha}^\dagger
b_{i \alpha}, \label{Ht}
\eeq
where the $\tau$ dependence of $Q_{ij}^v$ and $\lambda_i^v$ is chosen so that the vison executes the motion shown in Fig.~\ref{visonberry},
while always maintaining the constraint in Eq.~(\ref{const}).

We compute the Berry phase by working with the instantaneous ground state of $\mathcal{H}_b (\tau)$. This is facilitated by a diagonalization of the 
Hamiltonian by
performing a Bogoliubov transformation to a set of canonical Bose operators, $\gamma_{\mu\alpha}$, where the index $\mu = 1 \ldots N_s$,
where $N_s$ is the number of lattice sites. These are related to the $b_{i \alpha}$ by
\beq
b_{i \alpha} = \sum_\mu \left( U_{i \mu} (\tau) \gamma_{\mu \alpha} - V_{i \mu}^\ast (\tau)  \varepsilon_{\alpha\beta} \gamma^\dagger_{\mu\beta}\right) .
\eeq
The $N_s \times N_s$ matrices $U_{i \mu} (\tau), V_{i \mu} (\tau)$ perform the Bogoliubov transformation, and obey the following identities:\cite{sskag}
\beqn
\left( \begin{array}{cc} \lambda^v  & -Q^v  \\
-Q^{v\ast}  & - \lambda^v \end{array} \right) \left( \begin{array}{c} U \\ V \end{array} \right) &=& \hat{\omega} \left( \begin{array}{c} U \\ V \end{array} \right) \nn
U^\dagger U - V^\dagger V &=& 1 \nn
U U^\dagger - V^\ast V^T &=& 1 \nn
V^T U + U^T V &=& 0 \nn
U V^\dagger + V^\ast U^T &=& 0 ,
\label{ident}
\eeqn
where $\hat{\omega}$ is a diagonal matrix containing the excitation energies of the Bogoliubov quasiparticles,
and all quantities in Eq.~(\ref{ident}) have an implicit $\tau$ dependence.

We can use the above transformations to write down the instantaneous (unnormalized) wavefunction of the vison as the unique
state which obeys $\gamma_{\mu\alpha} \left| \Psi^v \right\rangle = 0$ for all $\mu$, $\alpha$:
\beq
\left| \Psi^v \right\rangle = \exp \left( \sum_{i<j} f^v_{ij} \,
\varepsilon_{\alpha\beta} b_{i \alpha}^\dagger b_{j \beta}^\dagger \right) |0 \rangle, \label{vpsi}
\eeq
where
\beq
f^v_{ij} = \sum_{\mu} \left( U^{-1 \dagger} \right)_{i \mu} \left( V^\dagger \right)_{\mu j}. \label{fij}
\eeq
Then the Berry phase accumulated during the $\tau$ variation of $\mathcal{H}_b (\tau)$ is 
\beq 
\frac{i}{\langle \Psi^v | \Psi^v \rangle} \mbox{Im}  \left\langle \Psi^v \right| \frac{d}{d \tau} \left| \Psi^v \right\rangle
= i \, \mbox{Im}\, \mbox{Tr} \left[ V^\dagger V \left( U^{-1} \frac{d U}{ d \tau} - V^{-1} \frac{dV}{d \tau} \right) \right]. \label{berry}
\eeq

We now assume that the Hamiltonian in Eq.~(\ref{Ht}) preserves time-reversal symmetry. Then, we can always choose a gauge in which
the $Q_{ij}$, $U_{i \mu}$ and $V_{i\mu}$ are all real. Under these conditions, the expression in Eq.~(\ref{berry}) vanishes
identically. It is clear that this argument generalizes to the case where we project the wavefunction to boson states which 
obey the constraint in Eq.~(\ref{const}).

We have now shown that no instantaneous Berry phase is accumulated during the vison motion of Fig.~\ref{visonberry}.
Under these conditions, the total gauge-invariant Berry phase is simply equal to the phase difference between
the wavefunctions in the initial and final states.\cite{rs2} As shown in Fig.~\ref{visonberry}, this phase difference is $\pi$.

\section{Effective Ising models for visons on various other lattice geometries}
\label{app:geom}

In this Appendix we summarize the results of a Ginzburg-Landau analysis of frustrated transverse field Ising models (TIMs) on  various lattice geometries. These models describe the low energy properties of different frustrated Heisenberg antiferromagnets in terms of their vison exciations. Some of these results have been discussed previously in the literature.

\subsection{Triangular lattice}

\begin{figure}
\begin{center}
\includegraphics[width=6.5cm]{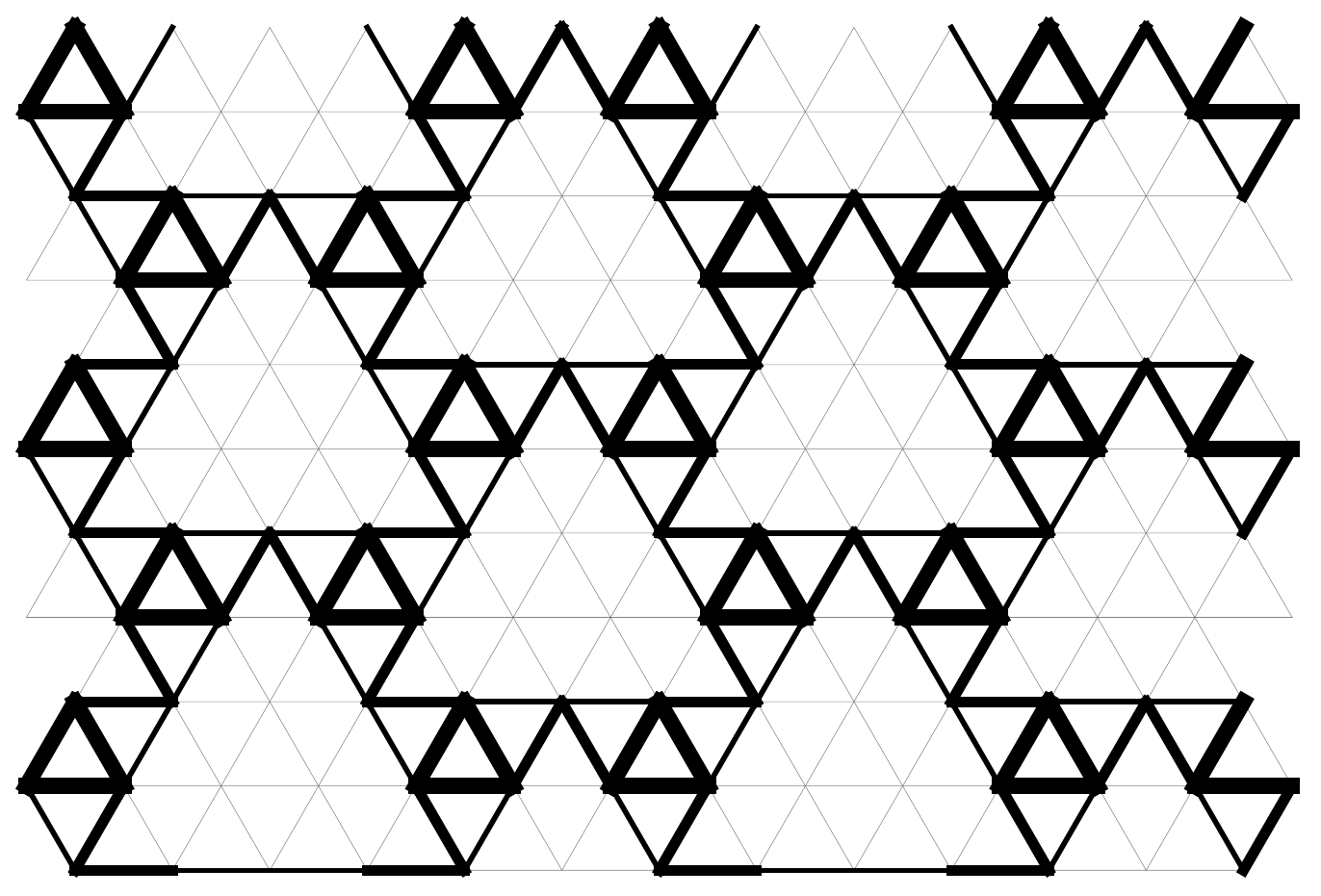}
\hspace{1cm}
\includegraphics[width=6.5cm]{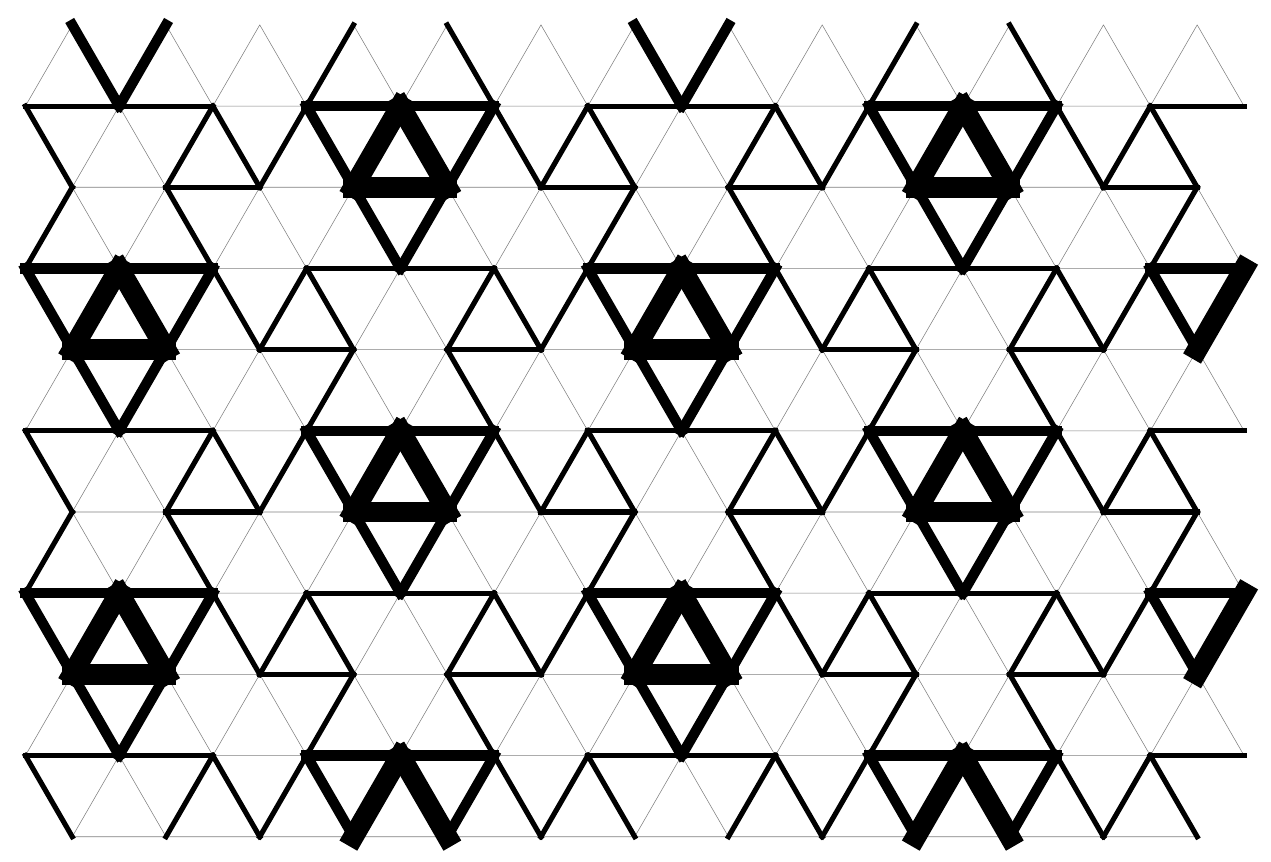}
\end{center}
\caption{Confining phase on the triangular lattice. Plotted is the gauge invariant bond-strength $J_{ij} \phi_i \phi_j$ for nearest neighbor bonds on the frustrated honeycomb lattice, shown on the corresponding triangular lattice bonds. Black lines represent satisfied bonds ($-J_{ij} \phi_i \phi_j < 0 $). There are no frustrated bonds in the confining phase. The thickness of the bonds is proportional to the bond-strength. The two different patterns arise due to a crossover when the sign of the O(4)-breaking term in the GL-functional changes.}
\label{fig:tri}
\end{figure}

The vison excitations of a Heisenberg antiferromagnet on the triangular lattice are described by a frustrated TIM on the dual honeycomb lattice, which has been studied previously by Moessner and Sondhi \cite{sondhi}. A Ginzburg-Landau analysis reveals four critical modes and the corresponding PSG transformation matrices generate a $288$ element subgroup of O(4) which is isomorphic to $(C_3 \times GL(2,\mathbb{Z}_3) ) \ltimes C_2$, \cite{gap} where $C_n$ denotes the cyclic group of order $n$. PSG matrices for a specific gauge choice can be found in Ref.\  \onlinecite{sondhi}. An O(4)-breaking term appears at sixth order in the GL-functional, the minimization of which gives rise to a single confined phase with a 24-site unit cell ({\em i.e.\/} a 12-site unit cell on the triangular lattice) that is symmetric under $2 \pi/3$-rotations and reflections. Bond patterns of this phase are shown in Fig.\ \ref{fig:tri}. Note that there is a transition when the sign of the O(4)-breaking term changes.

\subsection{Honeycomb lattice}

For the frustrated honeycomb lattice antiferromagnet the visons are described by an antiferromagnetic TIM on the dual triangular lattice. There are two critical modes at the Brillouin zone edges $\Q=\pm (4 \pi/3, 0)$ and the PSG matrices corresponding to translations $T$ by any basis vector, rotations $R_6$ and reflections $I_y$ about the y-axis of the two mode amplitudes are given by
\beq
T= -\big( \mathbbm{1}_2 + i \sqrt{3} \, \sigma_z \big)/2 \ , \hspace{2cm} R_6 = I_y = \sigma_x 
\eeq
where $\sigma_i$ denote the Pauli matrices. These PSG matrices generate the 6 element dihedral group $D_3$, {\em i.e.\/} the symmetry group of the equilateral triangle. The invariant Ginzburg-Landau functional has been discussed previously by Blankenschtein \emph{et al}.\ \cite{Blankenschtein}, who showed that an O(2) symmetry breaking term appears at sixth order.\cite{xu2} Minimizing the GL functional gives rise to only one possible confining phase which breaks the translational symmetry. For a particular sign of the sixth order term,\cite{xu2} the confining phase 
has a three-site unit cell ({\em i.e.\/} six sites per unit cell on the honeycomb lattice) and is symmetric both with respect to rotations and reflections; 
the corresponding dimer pattern on the honeycomb lattice has the maximal number of one perfectly flippable hexagon per six-site unit cell \cite{sondhi} and is identical to the VBS state found in Ref.~\onlinecite{rsb}. A plaquette-like phase is obtained for the other sign of the sixth-order term.\cite{xu2}
More complex minima structure for the vison dispersion have also been considered in Ref.~\onlinecite{xu2}.

\subsection{Square lattice}

The effective vison model for the frustrated square lattice Heisenberg antiferromagnet is a frustrated TIM on the dual square lattice.\cite{jalabert,balents} In an appropriate gauge the two critical modes appear at zero momentum\cite{footnote2} and the corresponding (gauge-dependent) PSG-matrices for translations $T_x$, rotations $R_4$ and reflections $I_x$ take the form
\beq
T_x= \big( \sigma_x + \sigma_z)/\sqrt{2} \ , \hspace{1cm} R_4 =  \sigma_z \ , \hspace{1cm} I_x = \mathbbm{1}_2 \ .
\eeq
These matrices generate the 16 element dihedral group $D_8$. An invariant GL-polynomial that breaks the O(2) symmetry appears only at eight order, as discussed by Blankenschtein \emph{et al}.\ \cite{Blankenschtein2}. Depending on the sign of this eight order term, two different confining phases are possible. Both phases are reflection symmetric, break the translational symmetries and have a four-site unit cell. One of the two phases is invariant under $\pi/2$-rotations, whereas the other one has a reduced rotational symmetry and is only invariant with respect to $\pi$-rotations. 
These are the familiar `plaquette' and `columnar' VBS states.\cite{dcp}

More complex vison dispersion structures, with further-neighbor couplings, have been described recently in Ref.~\onlinecite{xu2}.


\begin{thebibliography}{}
\bibitem{footnote1}A hint where to look for an interesting slice in the parameter space can be obtained by performing a fluctuation analysis of the frustrated transverse field Ising model \eqref{HIsing}. This calculation leads to the same GL-functional as the PSG analysis, but with a fixed set of couplings.
\bibitem{footnote2}Here we used a gauge where every second row of horizontal bonds is frustrated as well as an enlarged four-site unit cell, which ensures that the critical modes appear at zero momentum.
\end{thebibliography}
\end{document}